\documentclass[twocolumn]{aastex631}

\usepackage{amsmath}
\usepackage{array}   % for \newcolumntype macro
\usepackage{enumitem}
\usepackage{hyperref}
\begin{document}

\title{GOALS-JWST: Constraining the Emergence Timescale for Massive Star Clusters in NGC 3256}

\correspondingauthor{S. T. Linden}
\email{seanlinden@arizona.edu}

%Tentative author order
\author[0000-0002-1000-6081]{Sean T. Linden} 
\affiliation{Steward Observatory, University of Arizona, 933 N Cherry Avenue, Tucson, AZ 85721, USA}

\author[0000-0001-8490-6632]{Thomas Lai}
\affiliation{IPAC, California Institute of Technology, 1200 E. California Blvd., Pasadena, CA 91125}

\author[0000-0003-2638-1334]{Aaron S. Evans}
\affil{Department of Astronomy, University of Virginia, 530 McCormick Road, Charlottesville, VA 22904, USA}
\affiliation{National Radio Astronomy Observatory, 520 Edgemont Road, Charlottesville, VA 22903, USA}

\author[0000-0003-3498-2973]{Lee Armus}
\affiliation{IPAC, California Institute of Technology, 1200 E. California Blvd., Pasadena, CA 91125}

\author[0000-0003-3917-6460]{Kirsten L. Larson}
\affiliation{AURA for the European Space Agency (ESA), Space Telescope Science Institute, 3700 San Martin Drive, Baltimore, MD 21218, USA}

\author[0000-0002-5807-5078]{Jeffrey A. Rich}
\affiliation{The Observatories of the Carnegie Institution for Science, 813 Santa Barbara Street, Pasadena, CA 91101}

\author[0000-0002-1912-0024]{Vivian U}
\affiliation{Department of Physics and Astronomy, 4129 Frederick Reines Hall, University of California, Irvine, CA 92697, USA}

\author[0000-0003-3474-1125]{George C. Privon}
\affiliation{National Radio Astronomy Observatory, 520 Edgemont Road, Charlottesville, VA 22903, USA}
\affil{Department of Astronomy, University of Virginia, 530 McCormick Road, Charlottesville, VA 22904, USA}

\author[0000-0003-4268-0393]{Hanae Inami}
\affiliation{Hiroshima Astrophysical Science Center, Hiroshima University, 1-3-1 Kagamiyama, Higashi-Hiroshima, Hiroshima 739-8526, Japan}

\author[0000-0002-3139-3041]{Yiqing Song}
\affiliation{European Southern Observatory, Alonso de Córdova, 3107, Vitacura, Santiago, 763-0355, Chile}
\affiliation{Joint ALMA Observatory, Alonso de Córdova, 3107, Vitacura, Santiago, 763-0355, Chile}

\author[0000-0002-6570-9446]{Marina Bianchin}
\affiliation{Department of Physics and Astronomy, 4129 Frederick Reines Hall, University of California, Irvine, CA 92697, USA}

\author[0000-0002-4375-254X]{Thomas Bohn}
\affil{Hiroshima Astrophysical Science Center, Hiroshima University, 1-3-1 Kagamiyama, Higashi-Hiroshima, Hiroshima 739-8526, Japan}

\author[0009-0003-4835-2435]{Victorine A. Buiten}
\affiliation{Leiden Observatory, Leiden University, PO Box 9513, 2300 RA Leiden, The Netherlands}

\author[0000-0003-4286-4475]{Maria Sanchez-Garc\'ia}
\affiliation{Institute of Astrophysics, Foundation for Research and Technology-Hellas (FORTH), Heraklion, 70013, Greece}

\author[0000-0002-6650-3757]{Justin Kader}
\affiliation{Department of Physics and Astronomy, 4129 Frederick Reines Hall, University of California, Irvine, CA 92697, USA}

\author[0000-0003-4023-8657]{Laura Lenki\'c}
\affiliation{Stratospheric Observatory for Infrared Astronomy, NASA Ames Research Center, Mail Stop 204-14, Moffett Field, CA 94035, USA}
\affiliation{Jet Propulsion Laboratory, California Institute of Technology, 4800 Oak Grove Dr., Pasadena, CA 91109, USA}

\author[0000-0001-7421-2944]{Anne M. Medling}
\affiliation{Department of Physics \& Astronomy and Ritter Astrophysical Research Center, University of Toledo, Toledo, OH 43606, USA}
\affiliation{ARC Centre of Excellence for All Sky Astrophysics in 3 Dimensions (ASTRO 3D)}

\author[0000-0002-5666-7782]{Torsten B\"oker}
\affiliation{European Space Agency, Space Telescope Science Institute, Baltimore, Maryland, USA}

\author[0000-0003-0699-6083]{Tanio D\'iaz-Santos}
\affiliation{Institute of Astrophysics, Foundation for Research and Technology-Hellas (FORTH), Heraklion, 70013, Greece}
\affiliation{School of Sciences, European University Cyprus, Diogenes street, Engomi, 1516 Nicosia, Cyprus}

\author[0000-0002-2688-1956]{Vassilis Charmandaris}
\affiliation{Department of Physics, University of Crete, Heraklion, 71003, Greece}
\affiliation{Institute of Astrophysics, Foundation for Research and Technology-Hellas (FORTH), Heraklion, 70013, Greece}
\affiliation{School of Sciences, European University Cyprus, Diogenes street, Engomi, 1516 Nicosia, Cyprus}

\author[0000-0003-0057-8892]{Loreto Barcos-Mu\~noz}
\affiliation{National Radio Astronomy Observatory, 520 Edgemont Road, Charlottesville, VA 22903, USA}
\affiliation{Department of Astronomy, University of Virginia, 530 McCormick Road, Charlottesville, VA 22904, USA}

\author[0000-0001-5434-5942]{Paul van der Werf}
\affiliation{Leiden Observatory, Leiden University, PO Box 9513, 2300 RA Leiden, The Netherlands}

\author[0000-0002-2596-8531]{Sabrina Stierwalt}
\affiliation{Occidental College, Physics Department, 1600 Campus Road, Los Angeles, CA 90042}

\author[0000-0002-5828-7660]{Susanne Aalto}
\affiliation{Department of Space, Earth and Environment, Chalmers University of Technology, 412 96 Gothenburg, Sweden}

\author[0000-0002-7607-8766]{Philip Appleton}
\affiliation{IPAC, California Institute of Technology, 1200 E. California Blvd., Pasadena, CA 91125}

\author[0000-0003-4073-3236]{Christopher C. Hayward}
\affiliation{Center for Computational Astrophysics, Flatiron Institute, 162 Fifth Avenue, New York, NY 10010, USA}

\author[0000-0001-6028-8059]{Justin H. Howell}
\affiliation{IPAC, California Institute of Technology, 1200 E. California Blvd., Pasadena, CA 91125}

\author[0000-0001-6919-1237]{Matthew A. Malkan}
\affiliation{Department of Physics \& Astronomy, UCLA, Los Angeles, CA 90095-1547}

\author[0000-0002-8204-8619]{Joseph M. Mazzarella}
\affiliation{IPAC, California Institute of Technology, 1200 E. California Blvd., Pasadena, CA 91125}

\author[0000-0001-7089-7325]{Eric J. Murphy}
\affiliation{National Radio Astronomy Observatory, 520 Edgemont Road, Charlottesville, VA 22903, USA}

\author[0000-0001-7291-0087]{Jason Surace}
\affiliation{IPAC, California Institute of Technology, 1200 E. California Blvd., Pasadena, CA 91125}

%for which we perform aperture photometry
\begin{abstract}

%A direct comparison with archival HST imaging from 0.3-0.8$\mu$m reveals that $16 \%$ of these sources are undetected at optical wavelengths. 
%within LIRGs
% self-consistently
We present the results of a {\it James Webb Space Telescope} (JWST) NIRCam and NIRSpec investigation into the young massive star cluster (YMC) population of NGC 3256, the most cluster-rich luminous infrared galaxy (LIRG) in the Great Observatories All Sky LIRG Survey. We detect 3061 compact YMC candidates with a $S/N \geq 3$ at F150W, F200W, and F335M. Based on {\it yggdrasil} stellar population models, we identify 116/3061 sources with F150W - F200W $> 0.47$ and F200W - F355M $> -1.37$ colors suggesting they are young (t $\leq 5$ Myr), dusty ($A_{V} = 5 - 15$), and massive ($M_{\odot} > 10^{5}$). This increases the sample of dust-enshrouded YMCs detected in this system by an order of magnitude relative to previous HST studies.~With NIRSpec IFU pointings centered on the northern and southern nucleus, we extract the Pa$\alpha$ and 3.3$\mu$m PAH equivalent widths for 8 bright and isolated YMCs. Variations in both the F200W - F335M color and 3.3$\mu$m PAH emission with the Pa$\alpha$ line strength suggest a rapid dust clearing ($< 3 - 4$ Myr) for the emerging YMCs in the nuclei of NGC 3256. Finally, with both the age and dust emission accurately measured we use {\it yggdrasil} to derive the color excess (E(B - V)) for all 8 YMCs. We demonstrate that YMCs with strong 3.3$\mu$m PAH emission (F200W - F335M $> 0$) correspond to sources with E(B - V) $> 3$, which are typically missed in UV-optical studies. This underscores the importance of deep near-infrared imaging for finding and characterizing these very young and dust-embedded sources.

\end{abstract}


\section{Introduction}

Young massive star clusters (YMCs) form in the dense molecular gas cores within giant molecular clouds (GMCs) under high pressures and densities, and host the majority ($>80\%$) of the massive stars responsible for stellar feedback in galaxies. These objects are often found in merging galaxy systems, galactic nuclei, and blue compact dwarf (BCD) galaxies \citep[e.g.][]{bcm02b, bcm10, aa20a, linden23}. 

Despite advances made this last decade in understanding the formation and evolution of YMCs, properties such as the mass and physical size of the clouds from which they are formed are not well constrained. Numerical simulations have been utilized to propose that the feedback from YMCs depends on initial cloud conditions such as the GMC mass \citep[e.g.][]{dale14,howard17}, the surface density of the interstellar medium \citep[ISM;][]{kim18}, the metallicity \citep{fukushima20}, and the overall turbulence \citep{geen18, guszejnov22}. While there are still many uncertainties, studies suggest that photoionization may dominate over other pre-supernova (SNe) mechanisms such as stellar winds \citep{barnes21, geen21, ali22} and radiation pressure \citep{murray10}. These mechanisms set the structure into which SNe explode, potentially clearing low-density channels through which energy can escape \citep{lucas20,bending22}. However, we still do not understand how this process of star cluster emergence from their birth clouds depends on the properties of the surrounding ISM, or indeed the larger galactic environment.

By combining UV/optical photometry of star clusters with H$\alpha$ morphology of H{\sc ii} regions from the Hubble Space Telescope (HST), several authors have derived emergence timescales $< 4 - 5$ Myr, and as short as 2 Myr, in nearby spiral galaxies \citep{bcm11, hollyhead15, hannon19, hannon22}. However, these studies use UV and optical data, and are thus limited to the relatively dust-free components of star formation in normal star-forming galaxies. The addition of millimeter CO data to trace molecular clouds and, in some cases, 24$\mu$m imaging from the Spitzer Space Telescope to trace the dust-enshrouded star formation has enabled the use of both frequency and positional analysis to derive emergence timescales; \citet{matthews18}, \citet{grasha18}, \citet{grasha19}, \citet{kruijssen19}, \citet{chevance20a}, \citet{chevance22}, and \citet{bonne23} all conclude that timescales are short, only $3 - 5$ Myr, and likely shorter than the timescale for supernova explosions. 

More recently, \citet{kim23} used {\it James Webb Space Telescope} (JWST) mid-IR imaging, combined with CO imaging, of the nearby star–forming galaxy NGC 628 to infer that the embedded phase of star formation lasts about 5 Myr, during the first half of which dust obscuration is so high that the H$\alpha$ emission is not detectable. In a complementary fashion, \citet{whitmore23} combine HST optical with JWST near- and mid-IR medium and broad–band imaging of the galaxy NGC1365 to conclude that massive (M $\sim 10^{6} M_{\odot}$) star clusters in this galaxy remain completely or partially obscured for about 4 - 5 Myr. However, these studies do not have information on the near-infrared hydrogen recombination lines, which are key for constraining the ages of young star clusters when these are marginally detected or undetected at optical wavelengths. With the unprecedented sensitivity and resolving power of JWST we are now able to identify and characterize individual YMCs forming in the densest and dustiest regions of the ISM in starburst galaxies out to $\sim 100$ Mpc in the near-infrared (NIR).

In this paper we present JWST NIRCam and NIRSpec observations of the embedded YMC population in NGC 3256, the most cluster-rich luminous infrared galaxy (LIRG: defined as $L_{\rm IR} [8-1000\mu {\rm m}] \geq 10^{11.0}$ L$_\odot$) in the Great Observatories All Sky LIRG Survey (GOALS) JWST Early Release Science (ERS) campaign. LIRGs host the most extreme stellar nurseries in the local Universe. The activity in LIRGs is primarily triggered by gas-rich galaxy interactions, and at their peak, local LIRGs have star formation rates (SFRs) $\sim 100$ times higher than the Milky Way, placing them well above the star formation main sequence (SFMS) \citep{speagle14}. Many local LIRGs also have individual star-forming clumps with sizes of $50 - 100$pc and SFR surface densities ($\Sigma_{SFR}$) of $0.1 - 10$ $M_{\odot}$yr$^{-1}$kpc$^{-2}$) that rival those of high-$z$ galaxies \citep{klarson20}. 

Given that LIRGs have molecular gas surface densities an order of magnitude higher than nearby normal galaxies \citep{sg22,brunetti21,brunetti24}, the cluster emergence timescale may significantly deviate from what has been observed in previous studies. Such a difference would represent a fundamental shift in the relative roles of SNe, radiation pressure on dust grains from massive stars, and stellar wind feedback in starburst galaxies. Therefore, local LIRGs are one of the best laboratories for studying YMC formation and feedback in extreme environments at high-resolution. The observations described in this paper represent the most comprehensive census to date of embedded YMC formation and evolution in NGC 3256 in the NIR and the first constraints on the emergence timescale in a nearby LIRG.

Throughout this letter, we adopt a WMAP Cosmology of $H_0 = 69.3$ km s$^{-1}$ Mpc$^{-1}$, $\Omega _{\rm matter} = 0.286$, and $\Omega _{\Lambda} = 0.714$ \citep[e.g.,][]{wmap}.

%The blue and red cross marks denote the location of the northern and southern nuclei respectively.
\begin{figure*}[]
\centerline{\includegraphics[width=5.46truein]{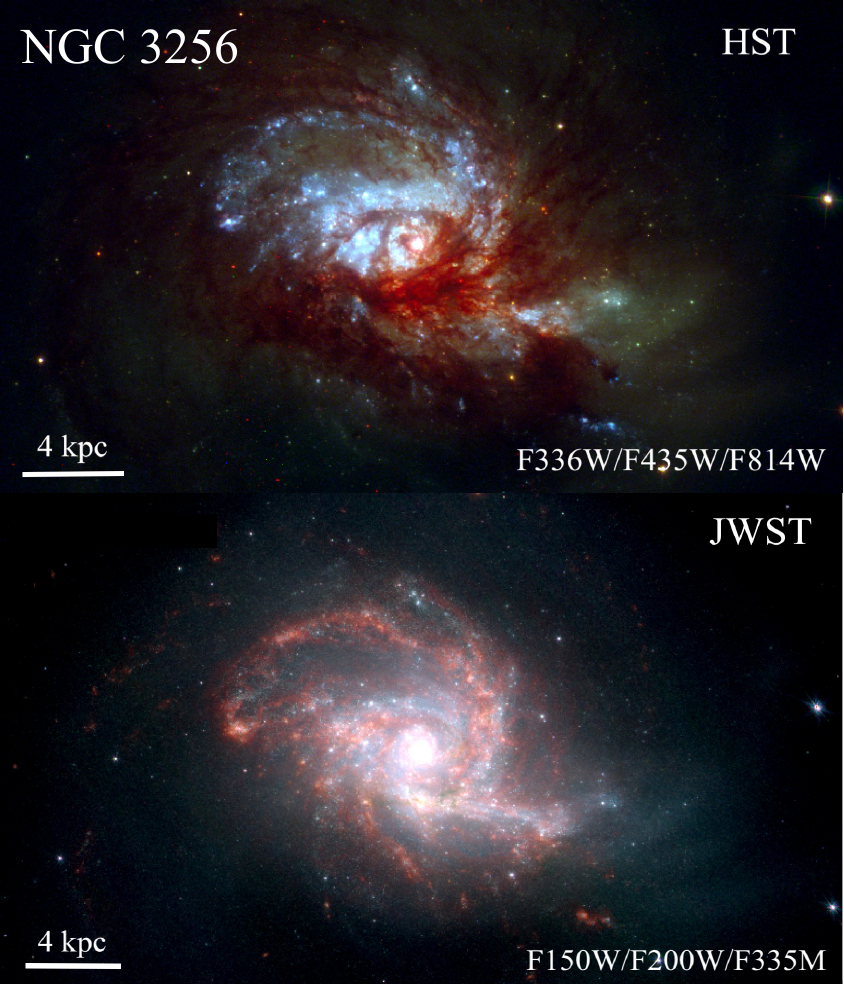}}
\caption{{\bf Top:} False-color HST imaging of NGC 3256 using F336W ($\lambda = 0.336 \mu$m), F435W ($\lambda = 0.435 \mu$m), and F814W ($\lambda = 0.814 \mu$m). The images are oriented N up E left, and the 4 kpc scale bar is equivalent to $18"$. Bright UV-optically visible clusters are seen predominantly in the dust-free regions of the northern galaxy. These clusters were identified and subsequently characterized in \citet{linden21}. {\bf Bottom:} False-color JWST NIRCam image of NGC 3256 using F150W, F200W, and F335M observations. The JWST data reveal many bright and red sources previously hidden behind the dust lanes obscuring the southern nucleus and parts of the northern galaxy where UV observations are not able to detect these embedded YMCs. The combination of the JWST filters used in this analysis enables us to characterize the properties of these red sources in NGC 3256 for the first time.}
\end{figure*}

\section{Observations}

JWST imaging and spectroscopy of NGC 3256 was obtained as part of the ERS program ``A JWST Study of the Starburst-AGN Connection in Merging LIRGs" (PID 1328; Co-PIs: Armus, Evans).

Near-Infrared Camera \citep[NIRCam;][]{nircam2} observations of NGC 3256 were taken on December 27th, 2022, and retrieved from the Mikulski Archive for Space Telescopes (MAST). The galaxy was imaged using the F150W ($\lambda=1.5\mu$m) and F200W ($\lambda=2.0\mu$m) short wavelength (SW) and F335M ($\lambda=3.35\mu$m) and F444W ($\lambda=4.4\mu$m) long wavelength (LW) filters for 644s each with module B. These observations utilized the `FULL' array with the `RAPID' readout mode and the `INTRAMODULE' 3-pt dither pattern. The raw data have been reduced using the JWST calwebb pipeline (CRDS 11.17.14, 1185.pmap). In our analysis we focus on F150W, F200W, and F335M to constrain the spectral energy distribution (SEDs) of individual star clusters and the amount of 3.3 $\mu$m PAH emission present within these sources. The level-3 products (.i2d) were then aligned to the Gaia DR2 reference frame using the Drizzlepac module TweakReg \citep{gaiadr2}. Finally, our F150W, F200W, and F335M science frames were drizzled to a common scale of 0.05"/pixel, which was chosen to match the angular resolution of existing HST WFC3 and ACS UV-optical observations of NGC 3256, and results in a physical resolution of 10 pc/pixel at the distance of NGC 3256 ($d_{L} = 44$ Mpc).

The JWST NIR spectroscopic observations of the NGC 3256 nuclei were obtained with NIRSpec \citep{jakobsen22} in its Integral-Field Unit (IFU) spectroscopy mode \citep{boker22} on December 23rd, 2022. The NIRSpec IFU observations were carried out using a set of high-resolution gratings (R$\sim2700$), including G140H/100LP, G235H/170LP, and G395H/290LP, covering the wavelength of 0.97–5.27 $\mu$m. For each grating, the exposure time was 467s and 1109s for the northern and southern nuclei, respectively. A 4-pt dither pattern was used to improve spatial resolution and to correct for bad pixels and cosmic rays in the extended star-forming regions around the nuclei. In this study we focus on the G325H and G395H spectra that contain the Pa$\alpha$ and 3.3 $\mu$m PAH emission lines respectively. The data reduction process was done using the JWST Science Calibration Pipeline version 1.12.5. Finally, `leakcal' images were obtained to mitigate the contamination due to failed-open shutters and overall leakage through the micro-shutter assembly (MSA).

In Figure 1 we compare HST optical and JWST NIR false-color images of NGC 3256. At optical wavelengths we see that NGC 3256 contains many bright clusters seen predominantly in the dust-free regions of the northern galaxy. These clusters are identified and subsequently characterized in \citet{linden21}. At longer wavelengths the JWST data reveal many bright and red sources previously hidden behind the dust lanes obscuring the southern nucleus (red cross) and parts of the northern galaxy where UV observations are not able to detect these embedded YMC candidates. Further, we see that strong 3.3um PAH emission, traced by the F335M filter, is present in and around the brightest young clusters and between the spiral arms in a filamentary morphology. This network of dust filaments is resolved for the first time in the NIR, and highlights the regions of dense gas and active star formation in NGC 3256. Finally, in the outer disk we see diffuse blue light in the NIR, indicative of an older stellar population of clusters and associations where very little gas and dust emission is seen.

\section{Results}

\subsection{Cluster Identification and Photometry}
 
Star cluster candidates in all three NIRCam filters were selected using the Source Extractor software \citep{sex}. We considered all sources with local S/N thresholds of $\geq 3$ in 2 or more contiguous pixels at F150W, F200W, and F335M utilizing 64 de-blending sub-thresholds, a background mesh size of 20 pixels, and a minimum contrast threshold of 0.0001. These detection parameters result in a candidate list of 4984 sources that are detected across all three filters.

We fit 2-D elliptical Gaussian profiles to all 4984 candidates to measure the major- and minor-axis FWHM at F150W. We then remove from our catalog all sources for which suitable fits could not be obtained (likely the result of multiple blended sources), as well as any source with a fitted major-axis FWHM that is $\geq 2$x larger than the FWHM of the F150W instrumental point-spread function (0.05" $= 10$ pc at the distance of NGC 3256) determined from WebbPSF \citep{perrin14}.~This requirement removes 1923 candidates, including 161 objects identified by-eye as either background galaxies or foreground stars. Here we focus on identifying compact star clusters, which can be best-modeled as single stellar populations (SSPs), and the derived properties can be compared directly to catalogs where strict size cuts have been applied. A full accounting of clumps and larger star-forming regions, which are likely a combination of multiple stellar populations, is outside the scope of this letter.

Photometry for 3061 compact clusters identified across the three NIRCam filters was then calculated using the IDL package APER. We used an aperture radius of 2 pixels ($0.1"$), and an annulus with an inner (outer) radius of 4 (6) pixels to measure the 3$\sigma$-clipped mean local background surrounding each cluster. Errors are estimated by varying the inner radius of the sky annulus from 2 - 5 pixels. Aperture corrections for F150W, F200W, and F335M were calculated based on the encircled energy values presented in \citet{gordon22}. We additionally applied a correction for foreground Galactic extinction, using the \citet{schlafly11} dust model and the empirical reddening law of \citet{fitz99}. We present our results using the Vega magnitude system for HST UV/optical photometric measurements and the AB magnitude system for all JWST NIRCam photometric measurements. The reason for this difference is to better-compare our UV/optical color-color diagram to those found in the literature (see Section 4.1).

% Regardless of the choice of SSP model, it is clear that 
\begin{figure*}[]
\centerline{\includegraphics[width=6.8truein]{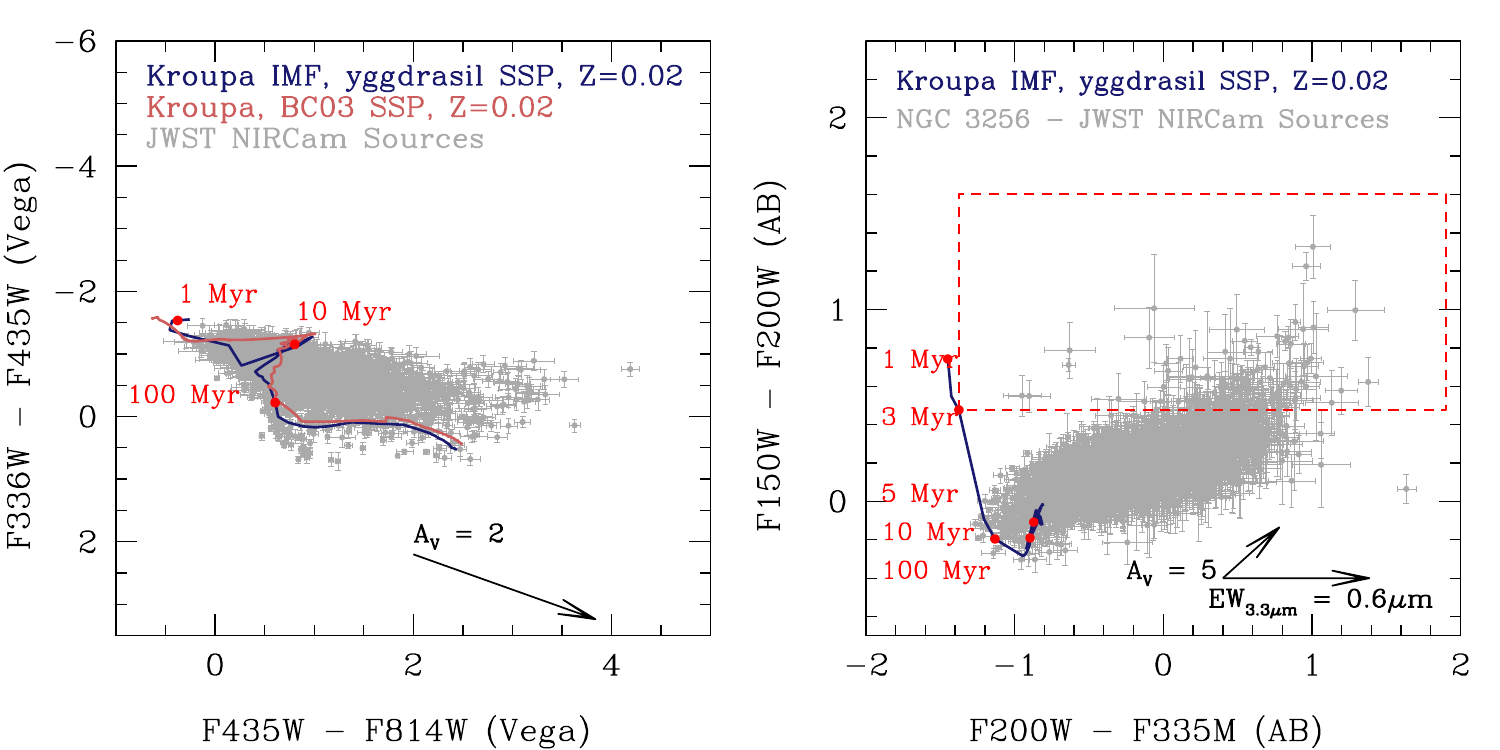}}
\caption{{\bf Left:} The F336W - F435W vs. F435W - F814W color-color diagram for compact YMC candidates with a $S/N > 3$ in all three HST UV/optical filters. In dark blue we overlay the {\it yggdrasil} SSP model tracks with solar metallicity and a Kroupa IMF. In red we overlay the equivalent models using the \citet{bc03} model tracks. Ages are marked along the sequence. Based on the extinction vector shown in the bottom-right, sources that are detected across all three HST filters are primarily YMC candidates with an $A_{V} < 2$. {\bf Right:} The F150W - F200W vs. F200W - F335M color-color diagram for all 3061 compact point-sources identified in our JWST NIRCam images (grey). Overlaid in dark blue is the {\it yggdrasil} SSP model track adopted here with solar metallicity and a Kroupa IMF. Ages are marked along the sequence. The bottom-right arrows represent an extinction of $A_{V} = 5$ and the maximum contribution to the F335M flux from 3.3$\mu$m PAH emission determined for star-forming regions within galaxies observed as part of the PHANGS-JWST survey respectively \citep{sandstrom23}. The red box represents the embedded YMC candidate selection presented in \citet{linden23}, where nearly all sources are identified in the JWST imaging alone. In NGC 3256 these candidates all appear to be young(t $\leq 5$ Myr) and heavy dust-enshrouded ($A_{V} = 5 - 15$).}
\end{figure*}

\subsection{Contributions to the Near-IR Colors of Young Massive Star Clusters}

%Several studies have searched for systematic variations between $f_{cov}$ and other ISM properties given that the most luminous regions, which form in the densest and dustiest regions of the ISM, may have lower escape fractions (and thus higher gas covering fractions) due to a higher proportion of the Lyman continuum photons being absorbed by dust \citep[e.g,][]{arthur04}.
For all confirmed clusters, the measured 3-band magnitudes are compared against SSP evolutionary models generated using the isochrone synthesis code {\it yggdrasil} \citep{yggdrasil}. This code computes the evolution of an instantaneous SF burst based on a Kroupa IMF and the Padova-AGB stellar evolution tracks over an age range of 1 Myr to 10 Gyr \citep[e.g.,][]{padova94}. Additionally, these models use CLOUDY \citep{cloudy17} to add the contribution from nebular emission lines by varying the fraction of ionizing photons that that interact with the surrounding gas ($f_{cov}$). \citet{teh23} explored the variation in the escape fraction for star clusters identified in the nearby galaxy NGC 628, finding average values of 0.1-0.2 ($f_{cov} \sim 80-90\%$) for the brightest regions, and no clear systematic trends with cluster age, mass, E(B - V), or galactocentric radius. We therefore adopt a gas covering fraction which ranges between 50 and $100\%$ to encapsulate the observed range found in nearby galaxies.

Further, we choose to adopt solar metallicity models for the entire galaxy, as suggested based on the relatively flat metallicity gradient observed in NGC 3256 \citep{jrich12}. Using optical spectroscopy of individual star clusters, \citet{trancho07} found that metallicity range in NGC 3256 was between 1 and 1.5 $Z_{\odot}$. \citet{goddard10} demonstrated that for super-solar SSP models up to 2.0 $Z_{\odot}$ the predominant effect on the derived physical parameters is that the true stellar mass is underestimated for a super-solar cluster population by up to a factor of $\sim 2$. For clusters identified near the center of NGC 3256 (see Section 3.4) our assumption of solar metallicity may artificially increase the derived stellar mass of our spectroscopic candidates, however these YMCs would have a mass $\geq 10^{5} M_{\odot}$, and are therefore still not subjected to the effects of stochastically. Our model choices allow us to make consistent comparisons with results from previous studies of YMCs in LIRGs \citep{stl17,aa20a} as well as the Legacy Extragalactic UV Survey (LEGUS) of nearby normal star-forming galaxies, which adopt the same {\it yggdrasil} models to determine cluster ages, masses, and extinctions \citep[e.g.,][]{ryon17,mm18a,cook19}.

%younger or older than 5 Myr
%, and the sharp vertical jump in NIR model colors
In the right panel of Figure 2 we see that the {\it yggdrasil} model tracks decrease vertically from 1 - 5 Myr, approximately perpendicular to the direction of the extinction vector show in the lower-right. The vertical drop in the SSP model occurs when massive stars evolve from blue supergiants into the red supergiant (RSG) evolutionary phase. RSGs are young, massive, luminous stars that rapidly exhaust their core hydrogen. The luminosities of RSGs peak at $\sim 1 \mu$m with absolute J-band magnitudes of $M_{J} = -8$ to $-11$, rivaling the integrated light of Milky Way globular clusters \citep{larsen11}. Thus RSGs, when present, can dominate the NIR flux of a YMC. Therefore the F150W - F200W NIR colors can be used as an absolute age indicator for YMCs due to the strong effects RSGs have on the overall continuum shape \citep{gazak13}.

Outside the ionization front, less energetic photons interact with the neutral gas and molecules, producing photo-dissociation regions \citep[PDRs;][]{hat99}{}. Within the PDRs, dust grains absorb and reprocess light from UV and optical photons produced by stars, which heat the surrounding gas via the photoelectric effect, and the gas then cools by line emission. In the very small grains regime we find the Polycyclic Aromatic Hydrocarbons (PAH). In this family of hydrocarbons, carbon atoms are organized in planar hexagonal rings and hydrogen atoms lie at the boundary of the rings \citep{tielens08}. A PAH molecule can absorb a UV photon, triggering vibrational de-excitation through IR emission \citep{leger89}. PAH spectral features arise at 3.3, 6.2, 7.7, 8.6, 11.3, 12.7, 16.4, and 17 $\mu$m and are linked to the different vibrational modes of their molecules \citep{allamandola98}. Their emission properties strongly depend on their sizes; the number of carbon atoms, the internal temperature \citep{leger89}, and the different ionization states of the PAH molecules \citep{dl07,dl21}. With JWST NIRCam observations at F335M we are able to image the 3.3$\mu$m PAH emission at $\sim 20$ pc resolution corresponding to the PDRs surrounding our YMC candidates and determine the contribution of this emission feature to the measured F335M flux \citep[e.g][]{dale23}.

\subsection{The Color-Color Diagram}

The distributions of star clusters in color-color diagrams have long been studied to gain insight into the properties and evolution of the cluster population \citep[e.g.][]{larsen00,rc10c,aa17}. In the left panel of Figure 2 we plot the HST F336W ($\lambda=0.336\mu$m), F435W ($\lambda=0.435\mu$m), and F814W ($\lambda=0.814\mu$m) color-color diagram for 2579/3061 compact YMC candidates identified with NIRCam that also have a $S/N > 3$ in all three HST UV/optical filters. Photometry was performed using the same apertures and background annuli adopted for the JWST NIRCam sources. In dark blue we overlay the {\it yggdrasil} SSP model tracks with solar metallicity and a Kroupa IMF. In Red we overlay the equivalent models using the \citet{bc03} model tracks. Regardless of the choice of SSP model, it is clear that requiring sources to be detected across all six HST and JWST filters, and in particular the F336W WFC3/UVIS filter, results in YMC candidates that are predominantly dust-free ($A_{V} < 2$). 

In the right panel of Figure 2 we plot the F150W - F200W vs. F200W - F335M color-color diagram for all 3061 compact point-sources identified in our JWST NIRCam images (grey). The red box represents the embedded YMC selection presented in \citet{linden23}, where nearly all sources are identified in the JWST imaging alone. Based on their location relative to the SSP models, these 116 YMC candidates all appear to be very young and heavily dust-enshrouded. Further, using the {\it yggdrasil} SSP models we take the average $M/L_{F200W}$ ratio from 1 - 3 Myr as well as the Solar AB absolute magnitude at F200W (\citet{willmer18}) to derive stellar masses of $10^{5.8-6.1} M_{\odot}$ for these sources. We note that the M/L ratio varies by less than a factor of 2 over this age range. 

\citet{sandstrom23} used F300M, F335M, and F360M observations as part of the PHANGS-JWST survey to accurately determine the continuum subtracted 3.3$\mu$m PAH emission across the disks of NGC 628, NGC 1365, and NGC 7496. They found that between 5\% and 65\% of the F335M intensity comes from the 3.3$\mu$m PAH feature within the inner regions of their targets. This percentage also systematically varies from galaxy to galaxy and shows radial trends related to each galaxy’s distribution of stellar mass, interstellar medium, and star formation rate. These results agree with \citet{lai20}, which shows the 3.3$\mu$m PAH on average contributes $\sim 15 - 20\%$ in the F335M filter in a sample of local LIRGs and ULIRGs (see Figure 17 in the paper). In the bottom-right we display a vector which represents the increase in F200W - F335M color assuming the maximum contribution (65$\%$) to the F335M flux from 3.3$\mu$m PAH emission within star-forming regions relative to a YMC without any PAH emission, as represented by the {\it yggdrasil} SSP model tracks. This vector demonstrates that the reddest sources detected with NIRCam are not simply shielded in large columns of dust, but themselves may have substantial dust emission that needs to be accounted for in order to use SSP models to derive accurate age, mass, and extinction values in the NIR. The combination of the extinction and emission vectors produce the elongated distribution of YMC candidates we see in the right panel of Figure 2.

\begin{figure}[]
\centerline{\includegraphics[width=3.5truein]{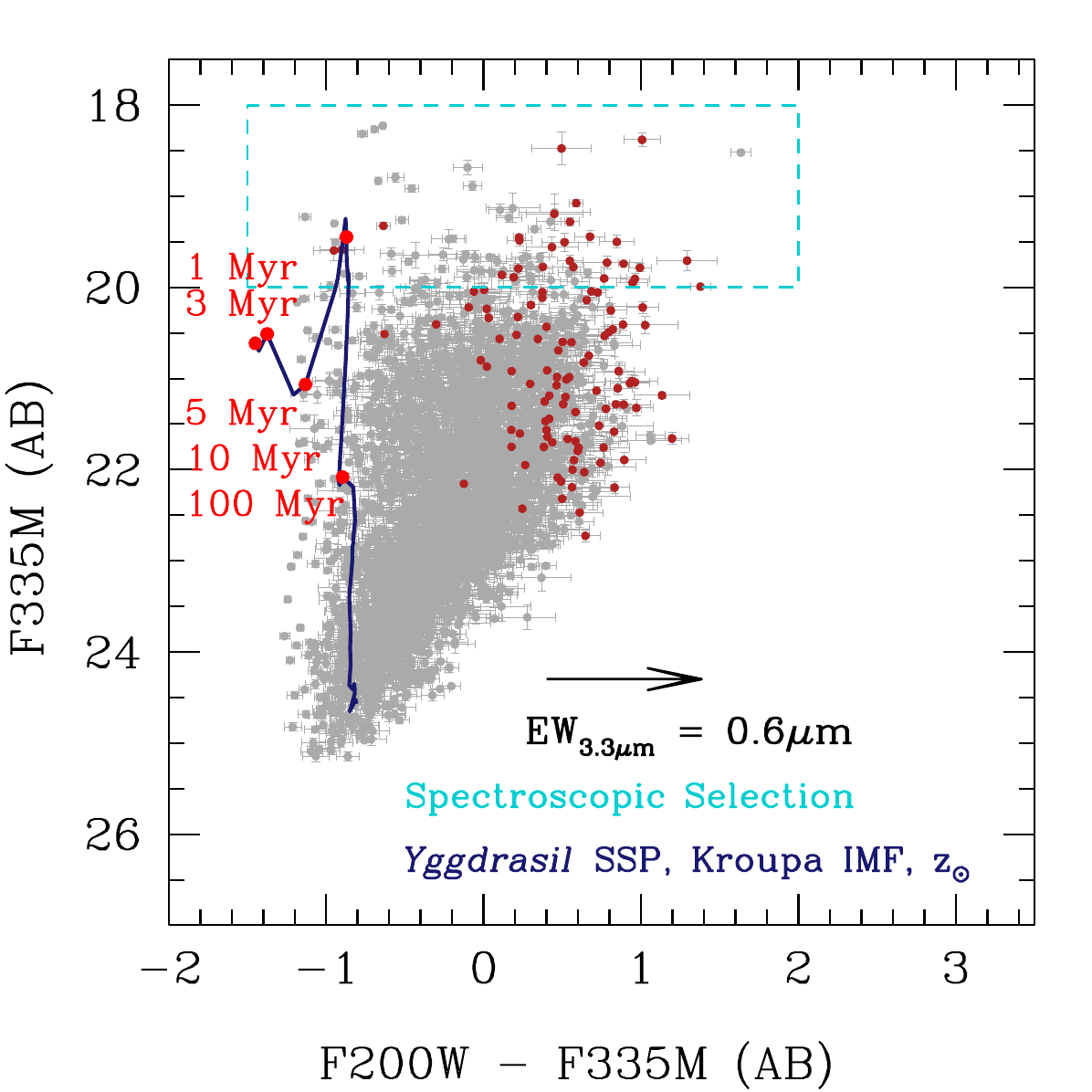}}
\caption{The F335M vs. F200W - F335M color-magnitude diagram with all 3061 compact YMCs detected shown in grey. In dark blue we plot the {\it yggdrasil} SSP model adopted in this analysis along with the same age markers as in the right panel of Figure 2. In dark red we highlight all 116 embedded YMC candidates based on their F150W - F200W and F200W - F335M colors. In turquoise we overlay the selection used to identify potential candidates within the NIRSpec FOV (3$''$ x 3$''$ $\sim 640$ x 640 pc), which have sufficient brightness to extract a high $S/N$ spectrum from 1 - 3 $\mu$m. The bottom-right arrow represents the maximum contribution from 3.3$\mu$m PAH emission to the F335M flux observed in \citep{sandstrom23}.}
\end{figure}

\begin{figure*}[]
\centerline{\includegraphics[width=5.0truein]{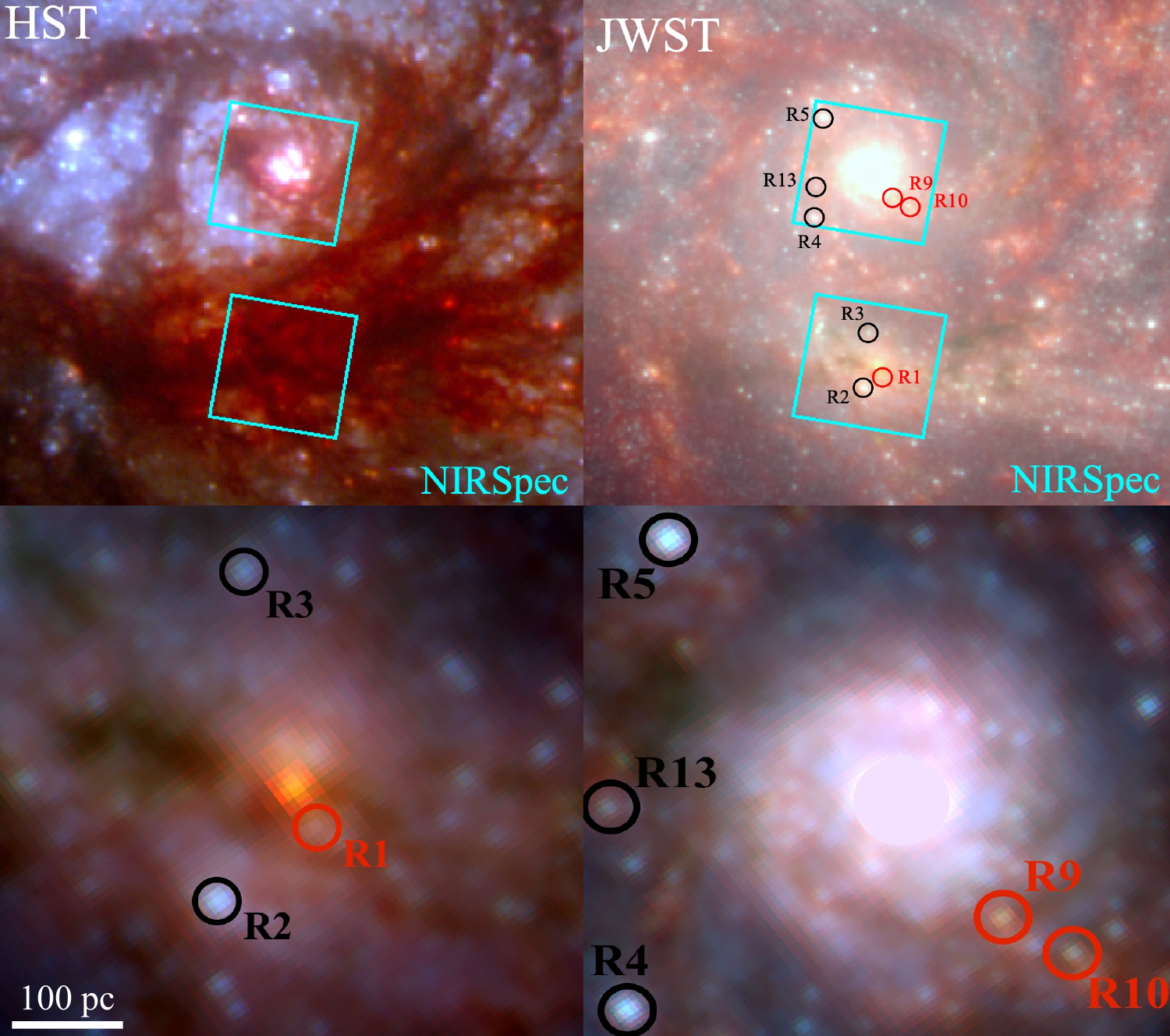}}
\caption{{\bf Top:} False color images of the northern and southern regions of NGC 3256 with both HST (F336W/F435W/F814W) and JWST NIRCam (F150W/F200W/F335M). In turquoise we overlay the NIRSpec FOV (3$''$ x 3$''$ $\sim 640$ x 640 pc) of the ERS observations used in this analysis. In black and red we overlay the apertures for the 8 YMC candidates within the selection box in Figure 3 that are sufficiently isolated such that an accurate local background can be determined for each cluster, and have sufficient $S/N$ to reliably extract both the Pa$\alpha$ and 3.3$\mu$m equivalent widths. {\bf Bottom:} NIRCam False-color zoom-in images of the FOV of the NIRSpec observations. Red circles represent candidates that fall within the embedded selection  based on their F150W - F200W and F200W - F335M colors (see Figures 2 and 3).}
\end{figure*}

\subsection{NIRSpec Spectroscopic Selection}

The identification of YMCs in local LIRGs has opened an unprecedented opportunity to determine the rapidly changing physical properties of the gas and dust surrounding massive star clusters as they form and evolve and eventually become optically-visible. In particular, the 1 - 5$\mu$m range carries key signatures that uniquely probe the physical conditions of the ionized gas and the spectral type of the most massive stars present, and hence, probe the strength of the radiation field in YMCs and the type of feedback (photoionization vs. mechanical) dominating in the different classes of objects \citep[e.g.][]{lumsden03,cresci10,bik12}. In particular, the ro-vibrational H$_{2}$ molecular emission lines in the 2 - 5$\mu$m range can be used to compare the physical properties of the hot molecular gas to the radiation field and ongoing feedback processes within photodissociation regions \citep[PDR -][]{hanson02,mn08,habart11}. NIRSpec observations allow us to correlate the relative strength of the 3.3$\mu$m PAH emission with the properties of the ionization field probed by ro-vibrational H$_{2}$ emission to determine how the physical properties of the dust and gas change as YMCs emerge from their parent GMCs.

%to our YMC candidates.
In Figure 3 we plot the F335M vs. F200W - F335M color-magnitude diagram for all 3061 compact YMCs detected shown in grey, and in dark red we highlight all 116 embedded YMC candidates based on their F150W - F200W and F200W - F335M colors selected in the right panel of Figure 2. Our selection method has identified embedded YMC candidates over 5 mag in apparent F335M brightness, demonstrating the depth of the NIRCam ERS observations. In turquoise we select the brightest sources at 3.3$\mu$m ($m_{F335M} < 20$) such that any potential YMC candidate identified within the NIRSpec FOV (3$''$ x 3$''$ $\sim 640$ x 640 pc) will have sufficient $S/N$ to extract spectrum of the Pa$\alpha$ emission line and the 3.3$\mu$m PAH feature. This requirement identifies 98 YMC candidates, and further requiring sources to be located within the NIRSpec footprint reduces our spectroscopic candidates from 98 to 13. From these 13 sources we remove a further 5, which blend multiple NIRCam-detected sources at the resolution of our combined channel 1 - 4 NIRSpec cubes. The spectra of the remaining 8 YMC candidates are then extracted using $0.2"$ apertures. For each spectrum we extract a local background in a nearby aperture with the same aperture size that appears to be emission free, and represents our best way to remove the contribution from the surrounding ISM.

%that are sufficiently isolated such that an accurate local background can be determined for each cluster, and have sufficient $S/N$ to reliably extract both the Pa$\alpha$ and 3.3$\mu$m PAH equivalent widths
%suggesting that these sources have already cleared the majority of the gas and dust surrounding them. 
In the top panels of Figure 4 we we show the 3-color UV/optical observations (F336W, F435W, F814W) with HST and the NIR observations (F150W, F200W, F335M) with JWST NIRCam of the northern and southern nuclear regions. In the top right panel we overlay our extraction apertures ($0.2"$) for our final 8 YMC candidates within the selection box presented in Figure 3. Red circles represent candidates that fall within the embedded selection presented in the right panel of Figures 2 and 3, demonstrating that within the NIRSpec footprint we have both red and blue YMC candidates for which we can extract accurate spectra. In the bottom panels of Figure 4 we see that 3 YMC candidates are visible in the HST observations (R4, R5, and R13) and are predominantly blue in the NIR. The two embedded candidates we extract in this footprint (R9 and R10) are found SW of the northern nucleus, and are seen as very red sources in the NIR imaging. In the southern footprint we see that previous HST observations completely miss any of the compact sources we identify with NIRCam imaging, including our blue YMC candidates (R2 and R3). Further, we find that the embedded YMC candidate in the southern footprint (R1) is found SW of the southern nucleus. Crucially, we see in the bottom panels of Figure 4 that our 8 YMC candidates are all sufficiently isolated such that no other compact point source is contained within the aperture used to extract NIRSpec spectra. These 8 sources represent the first NIRCam and NIRSpec joint analysis of YMCs in nearby galaxies to date. The photometric, spectroscopic, and SED-derived physical properties (including the age, mass, and extinction) for these YMC candidates are given in Table 1.

In the left panel of Figure 5 we show a spectral window (1.8-1.92$\mu$m) covering the Pa$\alpha$ recombination line extracted for all 8 YMCs ordered by their observed F200W - F335M color with the bluest source at the top and the reddest source at the bottom. We see that for the reddest YMC candidates multiple H$_{2}$ (1 - 0) transitions are found at 1.8358 (S(4)) and 1.892$\mu$m (S(5)) respectively. These lines are thought to trace hot molecular gas and the UV radiation field in the surrounding PDR of these YMCs \citep[e.g.][]{black87,wu19}.  We also see evidence for the He I (1.869$\mu$m) line which is transition only produced in star clusters hosting very massive stars \citep[$M_{\odot} > 100$][]{schreiber01,boker08}. The ionization energy of atomic helium is 24.6 eV, and thus is markedly higher than that of hydrogen (13.6 eV). Line emission from He I is therefore expected to arise predominantly in the vicinity of the most massive (and hence youngest) stars. Interpretation of the He I emission requires detailed photoionization models, and quantitative analysis of the temperature of the hottest stars in an H{\sc ii} region is subject to large uncertainties \citep{doherty95,lumsden01}. At least qualitatively, however, comparing the He I line strength to that of Pa$\alpha$ provides us a handle on the relative ages of the youngest stellar clusters because the hottest stars will vanish fastest \citep{boker08}. It is therefore notable that these diagnostics of massive stars and stellar feedback are only present within the reddest YMCs with the strongest Pa$\alpha$ line strengths in our NIRCam and NIRSpec cluster sample. 

%along with FeII at 1.810$\mu$m
\begin{figure*}[]
\centerline{\includegraphics[width=7.4truein]{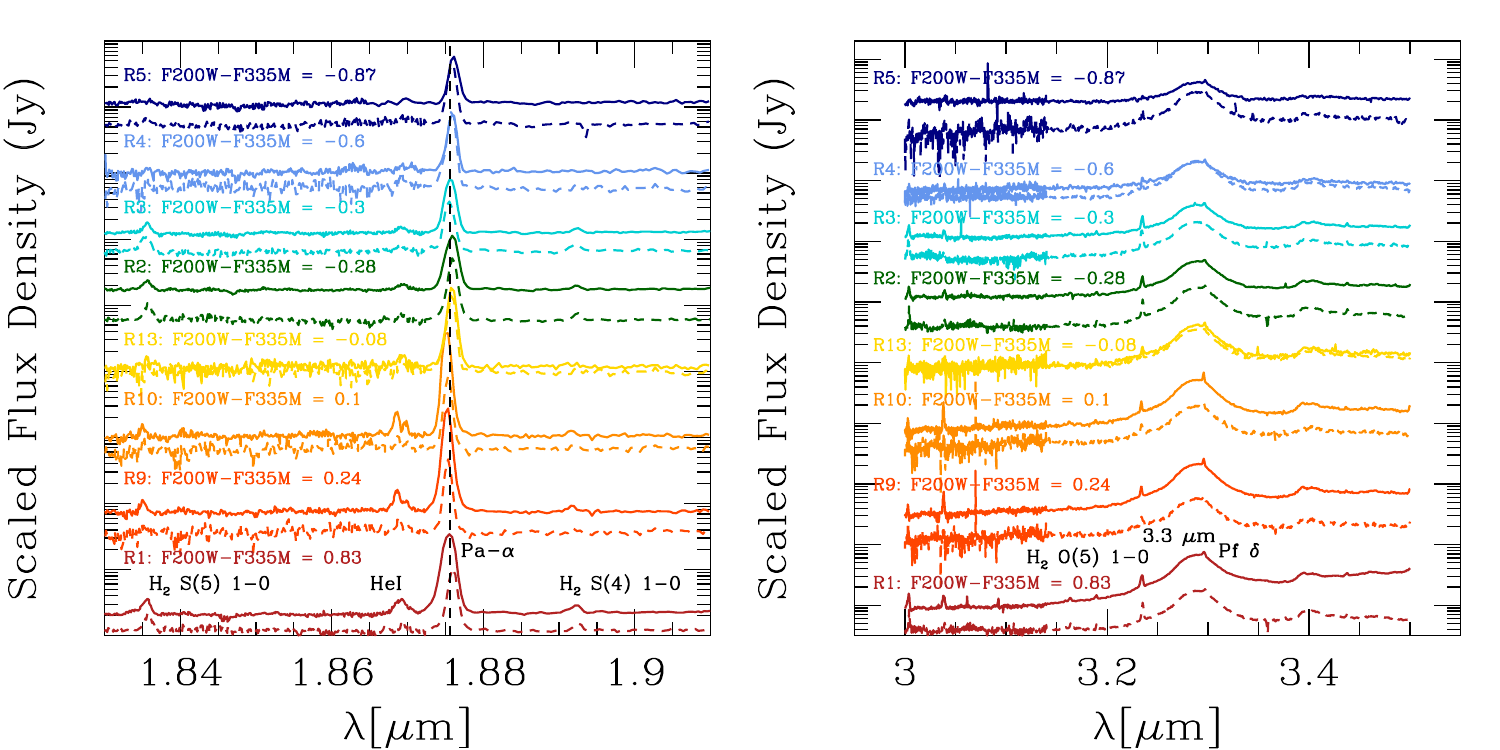}}
\caption{{\bf Left:}  A spectral window (1.8 - 1.92$\mu$m) showing the Pa$\alpha$ recombination line for all 8 YMCs ordered by their F200W - F335M color with the bluest source at the top and the reddest source at the bottom. The bottom 3 sources are classified as embedded YMCs based on our color criteria. We see that for these embedded YMCs multiple H$_{2}$ (1 - 0) transitions are found at 1.8358 (S(4)) and 1.892$\mu$m (S(5)) respectively, as well as He I at 1.869$\mu$m. The dashed lines represent the background spectra used to subtract the contributions of the surrounding ISM from each YMC spectra. {\bf Right:}  A spectral window (3 - 3.5$\mu$m) showing the 3.3$\mu$m PAH emission for the same 8 YMCs. We again see that for the reddest YMC candidates with the strongest PAH emission, the H$_{2}$ (1 - 0) transition at 3.2350 (O(5)) is easily detected along with the Pf$\delta$ recombination line at 3.2961 $\mu$m. The dashed lines represent the background spectra extracted for each source.}
\end{figure*}

% and the H$_{2}$ emission line, 
In the right panel of Figure 5 we show a spectral window (3 - 3.5$\mu$m) covering the 3.3$\mu$m PAH emission extracted for all 8 YMCs ordered by their observed F200W - F335M color with the bluest source at the top and the reddest source at the bottom. We again see that for the reddest YMC candidates an H$_{2}$ (1 - 0) transition is found at 3.2350 $\mu$m (O(5)) along with the Pf$\delta$ recombination line at 3.2961 $\mu$m. In conjunction with the correlations observed with the Pa$\alpha$ line strength, the 3.3$\mu$m PAH emission decreases as the NIR colors are observed to be bluer. 

This result is in broad agreement with \citet{lin20}, who use star cluster catalogs from LEGUS and 8 $\mu$m images from the IRAC camera on the Spitzer Space Telescope for five galaxies within 5 Mpc, to investigate how the PAH luminosity correlates with the stellar age on the 30-50 pc scale of star-forming regions. They find that star-forming regions in nearby galaxies show a tight anti-correlation between the 8$\mu$m dust-only luminosity and the age of primary stellar clusters younger than 1 Gyr. Further, \citet{egorov23} found a strong anti-correlation between the 3.3$\mu$m/7.7$\mu$m ratio, a proxy for the PAH abundance \citep{dl21}, and the ionization parameter in H{\sc ii} regions of nearby galaxies, confirming previous results that the survival of PAH molecules is connected to the properties of the radiation field and the evolutionary stage of the star-forming regions \citep[e.g.,][]{lai22,lai23}{}. Several scenarios could be responsible for the observed lack of PAH emission in star-forming regions. Coagulation onto dust grains might lead to the loss of a significant PAH fraction of $20-40\%$ in carbon atoms \citep{seok14}. Moreover, \citet{micelotta10} showed that interstellar shocks can also completely destroy the grains. These mechanisms play a crucial role in the evolution of PAHs, and they are strictly connected to the strength of the feedback as YMCs emerge from their birth clouds.

\subsection{Trends Between Derived Physical Parameters}

With a direct measurement of the background-subtracted Pa$\alpha$ and 3.3$\mu$m PAH equivalent widths (EWs) from the NIRSpec spectra, we can determine what the contribution is to the observed F200W - F335M colors for each of our 8 YMC candidates. In particular, we can derive the change in the NIRCam F335M magnitude ($F_{c}/(F_{c}+F_{l})= 2.512^{\Delta F335M}$) without the presence of the 3.3$\mu$m PAH emission using the EW and the bandwidth of the F335M filter (0.347$\mu$m). By removing this contribution we can then use the corrected F200W - F335M color to constrain the age and extinction (E(B - V)) with existing {\it yggdrasil} SSP models, which do not include contributions from dust emission. The final source of contamination to the measured 3.3$\mu$m PAH emission is the Pf$\delta$ recombination line at 3.296$\mu$m. In the right Panel of Figure 5 we see that for the reddest YMCs in our sample (R1, R9, and R10) a weak Pf$\delta$ line can be seen. \citet{lai20} estimated the average contribution of the Pf$\delta$ line to the integrated 3.3$\mu$m emission from their star-forming galaxy spectral template to be just $0.9\%$. This is in contrast to the starburst region SSC-N in II Zw 40, where correcting for the unusually strong recombination line results in a reduction of 3.3$\mu$m flux by $20.6\%$. However, the gas-phase oxygen abundance of SSC-N implies one-fourth to one-fifth solar metallicity (Z $\sim 0.004$) for II Zw 40 \citep{leitherer18}, which is much lower than any of the YMC candidates observed in NGC 3256, and therefore such strong contributions to the 3.3$\mu$m recombination are not seen.

In Figure 6 we show the 8 YMCs selected within our NIRSpec FOV as square symbols color-coded as in Figure 5. The vector attached to each point represents the adjusted F200W - F335M color taking into account the measured 3.3 $\mu$m PAH strength such that only the stellar emission associated with each source remains. We stress that although our selection was limited to the nuclear NIRSpec pointings available in the GOALS-ERS program, we are able to select YMC candidates that span nearly an order of magnitude in 3.3 $\mu$m PAH strength, and the full range of observed F200W - F335M colors in NGC 3256. With adjusted F200W - F335M colors for these YMCs we can use a standard $\chi^{2}$-minimization technique to find the best-fit age and E(B - V) values from a grid of {\it yggdrasil} SSPs using the model F150W, F200W, and F335M NIRCam fluxes. The resulting photometrically-derived ages all agree within 1$\sigma$ of our spectroscopic values derived using the Pa$\alpha$ equivalent width vs age models presented in \citet{sb99}. The $\chi^{2}$ values for each fit are given in Table 1. This demonstrates the effectiveness of our approach once the F200W - F335M color is properly corrected for the presence of 3.3 $\mu$m PAH emission. In the remainder of this section we present the results of these fits as well as the observed correlations between YMC age, NIR color, 3.3$\mu$m PAH strength, and color-excess E(B - V).

In the left panel of Figure 7 we plot the Pa$\alpha$ equivalent width-derived ages vs the observed F200W - F335M NIRCam color. We see a clear trend whereby the median age of YMC candidate within the embedded selection (F200W - F335M $\geq  0$) is 2 Myr, whereas blue YMCs (F200W - F335M $< 0$) have a median age of $> 5$ Myr. In the middle panel we see another clear trend where the YMCs with the strongest 3.3$\mu$m PAH emission are also the youngest. Our observations also suggest that 3.3$\mu$m PAH strength may be used as a proxy for cluster age out to 8 - 9 Myr in NGC 3256. Like in the left panel, we also see a clear trend when examining the Pa$\alpha$ equivalent width-derived ages vs the SED-derived color excess E(B - V), where the most heavily-extincted sources (E(B - V) $> 3$) have the youngest ages ($<$ 4 Myr). We find the reddest source within our NIRSpec sample, which is also the reddest source identified across the entire galaxy, has an E(B - V) $= 5.75$. This value is consistent with detailed SED models of the NIR-FIR spectrum of embedded YMCs presented in \citet{whelan11}, which demonstrated that for a star formation efficiency of $50\%$ and an outer envelope size of 25pc, visual extinctions as high as 18.3 mag should be expected even for a PDR clumpy fraction of $90\%$. Although rare, the detection of such YMCs demonstrates that the nuclear regions of local LIRGs are potentially the best laboratory for finding these extreme dust-buried systems with high SFEs that are likely to be progenitors of massive globular clusters in the future. 

\begin{figure}[]
\centerline{\includegraphics[width=3.5truein]{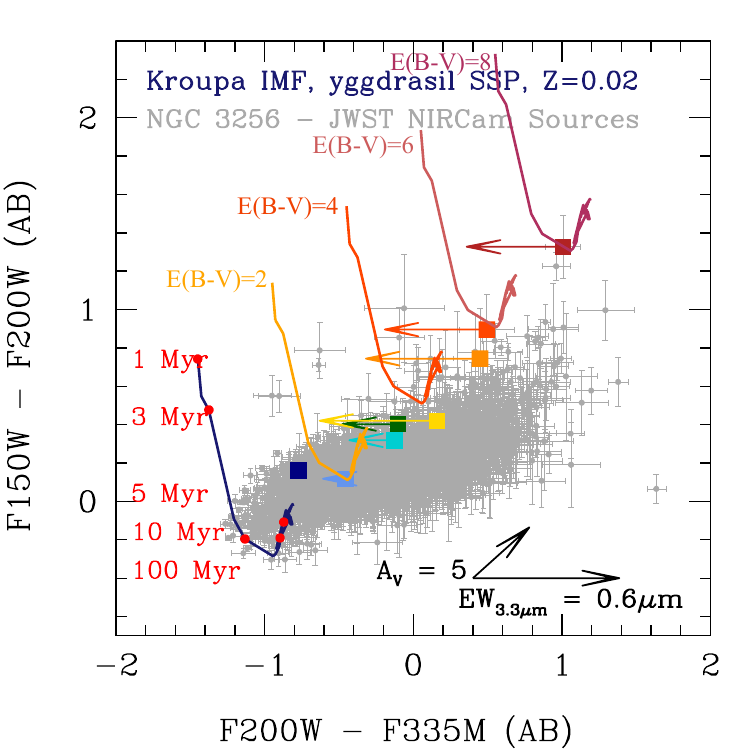}}
\caption{The F150W - F200W vs. F200W - F335M color-color diagram for all 3061 compact point-sources identified in our JWST NIRCam images (grey). Overlaid in dark blue to magenta are {\it yggdrasil} SSP model tracks with E(B - V) = 0, 2, 4, 6, and 8 respectively. The square points correspond to the 8 YMCs selected within our NIRSpec FOV color-coded as in Figure 4. The arrow attached to each point represents the adjusted F200W - F335M color taking into account the measured 3.3 $\mu$m PAH equivalent width such that only the stellar emission associated with each source remains. The bottom-right arrows represent an extinction of $A_{V} = 5$ and the maximum contribution from 3.3$\mu$m PAH emission to the F335M flux determined for star-forming regions within galaxies observed as part of the PHANGS-JWST survey, respectively \citep{sandstrom23}. It is clear that our NIRCam and NIRSpec YMC candidates span nearly an order of magnitude in 3.3 $\mu$m PAH strength, and the full range of observed F200W - F335M colors in NGC 3256}
\end{figure}

\begin{figure*}[]
\centerline{\includegraphics[width=7.4truein]{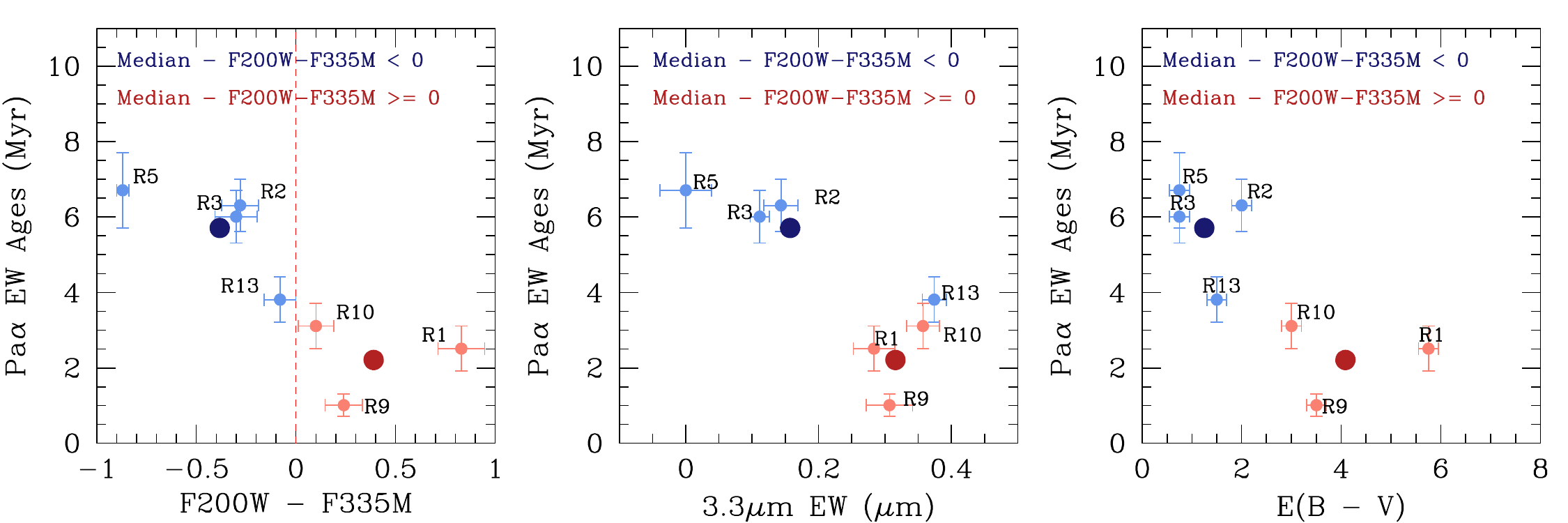}}
\caption{{\bf Left:} The Pa$\alpha$ equivalent width-derived age vs. the measured F200W - F335M color. We see a clear trend where the reddest YMCs in our sample have the youngest ages as determined from their recombination line equivalent widths. Based on the embedded selection presented in section 3.3, YMCs in NGC 3256 with F200W - F335M $> 0$ have ages $\leq 4$ Myr. {\bf Middle:} The Pa$\alpha$ equivalent width-derived age vs. the 3.3$\mu$m PAH equivalent width. We see a clear trend where the YMCs with the strongest PAH emission are also the youngest. {\bf Right:} The Pa$\alpha$ equivalent width-derived age vs. the SED-derived color excess E(B - V). Like in the left panel, we see a clear trend where the most heavily-extincted sources (E(B - V) $> 3$) have the youngest ages ($<$ 4 Myr).}
\end{figure*}

\section{Discussion}

\subsection{YMC photometric comparisons}

\citet{maschmann24} use HST F336W, F435W, F555W, and F814W observations of 38 galaxies as part of the PHANGS-HST survey to classify the color-color diagram into three distinct regions: the young cluster locus (YCL: $-1.5 <$ F336W - F435W $< -1$), the middle-age plume (MAP: $-1 <$ F336W - F435W $< 0$), and the old globular cluster clump (OGC: $0 <$ F336W - F435W $< 0.5$). This classification reveals that there is no correlation between the main sequence offset ($\Delta$MS) and the fraction of clusters in the YCL. The absence of a strong YCL feature above the MS is generally due to dust reddening and does not necessarily imply the absence of cluster formation. Further there is a strong linear correlation between a galaxy’s offset from the MS and the fraction of its cluster population in the MAP. Above the MS, the presence of a strong MAP feature indicates the elevated star and cluster formation activity must have been ongoing for about 100 Myrs. In the left panel of Figure 2 we see that sources detected with JWST NIRCam observations span both the YCL and MAP regions, allowing us to recover the full cluster formation history of the merger over the last $\sim 100$ Myr in NGC 3256.

% used INTEGRAL IFU observations to identify embedded star-forming regions in the center of the starburst LIRG Arp 299 based on recombination line ratios. 
With archival HST WFPC2 and NICMOS imaging \citet{garcia-marin06} found that the youngest sources within Arp 299, with ages of $\leq 3$ Myr, are the only ones with measured F160W - F222M colors $> 1$, suggesting that the youngest and most deeply embedded YMCs within nearby LIRGs display the reddest NIR colors. If indeed all of our reddest YMC candidates were young and embedded, it would increase the number of such sources identified previously through HST imaging alone by over an order of magnitude \citep{linden21}.

\citet{rodriguez22} present JWST NIRCam observations of YMC candidates in NGC 7496 as part of the PHANGS-JWST Cycle 1 Treasury Survey. They find that sources selected as having strong 3.3$\mu$m PAH emission based on the F300M - F335M color excess are predominantly young and embedded star clusters. With the corresponding CO (2 - 1) maps from the PHANGS-ALMA survey they correlate the locations of these embedded sources with peaks in the molecular gas emission suggesting that these sources have yet to significantly clear the gas and dust in their surrounding GMC. These results are consistent with our picture of rapid cluster emergence in a more extreme starburst system using NIRCam imaging of the northern and southern nuclear regions of NGC 3256.

\subsection{YMC spectroscopic comparisons}

%Unlike other star-forming regions identified in \citet{donnan24}, SF5 shows the highest obscuration and dust temperature as well as deep near-IR ices.
To compare our results with others in the literature we note that R9 in our spectroscopic YMC sample is also the SF5 region in \citet{donnan24}. These authors suggest that SF5 is a very dusty star-forming region with hot gas and dust around it, consistent with our findings that R9 is one of the youngest and reddest sources in our NIRCam and NIRSpec catalogs. We also note that although multiple supernovae have been detected in NGC 3256 \citep[e.g.][]{kankare18a,kankare18b}, none have been found at the location of SF5/R9. Further, \citet{galliano08} used NIR and MIR spectroscopy with the VLT to identify 3 very young (2 - 4 Myr), dust-enshrouded (A$_{V}$ = 3 - 13), and massive ($\sim 10^{6} M_{\odot}$) stellar clusters (M4, 5, and 6) within the central region of the nearby LIRG NGC 1365. In particular, M5 is identified in \citet{whitmore23} as the source with the brightest F335M magnitude and M6 has one of the reddest F300M - F335M colors, suggesting that strong PAH emission is indeed associated with the very youngest, and still embedded, star clusters. These results are consistent with the YMCs observed in NGC 3256.

% and in particular the H$_{2}$ (1 - 0) line flux decreases nearly 3 orders of magnitude between
\citet{herrera17} used ALMA and VLT observations of the overlap region of the Antennae galaxy to study how feedback in YMCs affects the surrounding ISM in a starburst event. They similarly find that both the Br$\gamma$ recombination line emission and the H$_{2}$ (1 - 0) S(1) emission decrease with increasing cluster age for YMCs with ages between 1 and 4 Myr. These observations suggest that most of the parent cloud has already been blown away, accelerated at the early stages of YMC evolution by radiation pressure, in a timescale of $\sim 1 - 2$ Myr in the overlap region of the Antennae.  Further, with VLT K-band spectroscopy of young stellar objects embedded within ultracompact H{\sc ii} regions identified in the Milky Way, \citet{bik06} determined that the reddest sources in their sample (with J - K colors $> 5$) have the strongest observed Br$\gamma$ recombination line fluxes, suggesting they have ages of $\sim 1$ Myr. 

These studies demonstrate that youngest and most deeply-embedded stars and star clusters exhibit both strong 3.3$\mu$m PAH emission and hydrogen recombination line emission, and that these trends appear ubiquitous in both normal and extreme starburst galaxies in the local Universe.

\subsection{The Emergence Timescale}

%in the nuclei of NGC 3256
Based on the trends observed in Figure 7, we conclude that YMCs in the nuclear regions of NGC 3256 appear to clear their surrounding dust quickly (E(B - V) $\geq 1$ by 4 Myr). A emergence timescale of $\sim 3$ Myr for massive YMCs strongly suggests that pre-supernova eedback remains an important mechanism to clear the surrounding gas and dust in these YMCs. \citet{krumholz19} suggest that for cluster masses $M > 10^{5} M_{\odot}$ the dominant mechanism responsible for this clearing is direct radiation pressure.

%, thus before SN explosions become significant,
%Large Magellanic Cloud, powered by one of the closest YMCs \citep{andersen09}, to compare the relative values of the pressures associated with radiation, ionized gas, and hot plasma, as a function of position in the nebula. 
Observations of 30 Doradus suggest that the cluster has broken apart its parent cloud within a timescale shorter than the main sequence life time of the most massive stars. Further, they find that within the shells close to the core of the central R136 cluster, radiation pressure dominates \citep{andersen09,lopez11} . Until now, it has not been possible to obtain complete samples of embedded YMCs beyond the Local Group to constrain feedback more directly for star-forming regions at higher metallicity, gas surface density, and ISM pressures, as the angular resolution of previous infrared facilities could not resolve the parsec scales required to identify YMCs at larger distances. 

Pre–SN feedback processes that are likely to be important for young star clusters in our mass and size range ($\sim 10^{5} M_{\odot}$ and $r \sim 10$ pc) include photoionization, direct radiation pressure and stellar winds \citep[e.g.][]{pellegrini11,dale12,krause13}, all of which have timescales of a few Myr, and should help clear the surrounding medium before the first supernova explosion occurs at around $\sim 4$ Myr \citep{leitherer14}.  However, as cluster mass increases early feedback mechanisms like stellar winds become increasingly ineffective, requiring SNe to fully impact and disperse the surrounding GMC, which may result in a longer emergence timescale ($\sim4 - 5$ Myr).

To further determine what early feedback mechanisms are important for our NIRCam-selected sources in NGC 3256, we follow the calculations presented in \cite{lopez14} to determine the radiation pressure in the H{\sc ii} regions surrounding our embedded YMC candidates. We adopt an average E(B-V) of 4 (see Figure 7), and a median Pa$\alpha$ flux of $5$x$10^{-16}$erg s$^{-1}$ cm$^{-2}$ Hz$^{-1}$. We further adopt a median cluster radius of 15 pc, to derive a value for the direct radiation pressure $P_{dir}$ of $5$x$10^{-9}$ g cm$^{-1}$ s$^{-2}$. This value is comparable to, and even slightly larger than the typical ionization ($P_{ion}$) pressures found for H{\sc ii} regions in the LMC of $\sim 1 - 5$x$10^{-10}$g cm$^{-1}$ s$^{-2}$ \citep{lopez14}. Further, our value is consistent with the results presented in \citet{della-bruna22}, which used MUSE to investigate the direct and ionization pressures for optically-selected H{\sc ii} regions in the disk and central regions of M83. They find that for the dustiest H{\sc ii} regions, $P_{dir}$ is approximately constant with galactocentric radius, and becomes comparable in strength to $P_{ion}$ in the central region. Therefore we conclude that direct radiation pressure must play an important role in the early feedback process for YMCs in the central region of NGC 3256.

We note that \citet{westmonquette14} analyzed three YMCs in the nearby starburst galaxy M82 that appear to have stalled feedback due to the high external pressure of the ISM in the central region. Indeed, the spectroscopic ages of those three clusters (4.5 - 6.4 Myr with E(B - V) $\sim 1.4 - 1.9$ mag), are similar to those of the bluer YMCs found in this work. On the other hand, \citet{levy21} find P Cygni profiles associated with three compact YMCs in the central starburst of NGC 253 suggesting they are actively expelling their surrounding gas and dust. Overall it is still unclear if YMCs are effective in rapidly clearing their surrounding dust and gas in all starburst environments, a question which requires a more complete survey of YMC formation and evolution in the centers of starburst galaxies to fully address.

\section{Summary}

We have presented the results of a {\it James Webb Space Telescope} NIRCam and NIRSpec investigation into the young massive star cluster population in the luminous infrared galaxy NGC 3256. We identify 3061 compact YMC candidates with a $S/N \geq 3$ at F150W, F200W, and F335M respectively for which we perform aperture photometry. Based on the derived NIR magnitudes and colors of our new YMC catalog we reach the following conclusions:

\vspace{0.2cm}
\noindent
(1). A direct comparison with archival HST imaging from 0.3 - 0.8$\mu$m reveals that $16 \%$ of these sources are undetected at optical wavelengths. Based on {\it yggdrasil} stellar population models, we identify 116 YMC candidates in our JWST imaging with F150W - F200W $> 0.47$ and F200W - F355M $> -1.37$ colors suggesting they are young (t $\leq 5$ Myr), dusty ($A_{V} = 5 - 15$), and massive ($M_{\odot} > 10^{5}$). This increases the sample of dust-enshrouded YMCs detected in NGC 3256 by an order of magnitude.

\vspace{0.2cm}
\noindent
(2).  With NIRSpec IFU pointings centered on the northern and southern nucleus, we extract the Pa$\alpha$ and 3.3$\mu$m PAH equivalent widths for 8 bright and isolated YMCs, that span nearly the full range of observed F200W - F335M colors in NGC 3356. Variations in both the F200W - F335M color and 3.3$\mu$m PAH line strength with Pa$\alpha$ equivalent width suggest a rapid evolutionary sequence ($< 3 - 4$ Myr) for photo-dissociation regions (PDRs) as the emerging YMCs clear their surrounding dust in the nuclei of NGC 3256 before supernova feedback can take effect.

\vspace{0.2cm}
\noindent
(3). With both the age and dust emission accurately measured we use {\it yggdrasil} to self-consistently derive the color excess (E (B -V)), demonstrating that YMCs with F200W - F335M $> 0$ correspond to sources with an E(B - V) $> 3$, which are completely missed in existing UV-optical studies, and underscores the importance of JWST for finding and characterizing these very young and dust-embedded sources within nearby galaxies.

\begin{acknowledgements}

The JWST data presented in this article were obtained from the Mikulski Archive for Space Telescopes (MAST) at the Space Telescope Science Institute. The specific observations analyzed can be accessed via \dataset[DOI]{https://doi.org/10.17909/7wqw-2r55}. S.T.L was partially supported through NASA grant HST-GO16914. A.S.E. was supported by NSF grant AST 1816838 and by NASA through grants HST-GO10592.01-A, HST-GO11196.01-A, and HST-GO13364 from the Space Telescope Science Institute, which is operated by the Association of Universities for Research in Astronomy, Inc., under NASA contract NAS5- 26555. V.U acknowledges funding support from JWST-GO-01717.001-A, HST-AR-17065.005-A, HST-GO-17285.001-A, and NASA ADAP grant 80NSSC23K0750.  H.I. and T.B. acknowledge support from JSPS KAKENHI Grant Number JP21H01129 and the Ito Foundation for Promotion of Science.~F. K. acknowledges support from the Spanish program Unidad de Excelencia Maria de Maeztu CEX2020-001058-M, financed by MCIN/AEI/10.13039/501100011033. AMM acknowledges support from the NASA Astrophysics Data Analysis Program (ADAP) grant number 80NSSC23K0750. This work was also partly supported by the Spanish program Unidad de Excelencia Mari\'a de Maeztu CEX2020-001058-M, financed by MCIN/AEI/10.13039/501100011033. This research has also made use of the NASA/IPAC Extragalactic Database (NED) which is operated by the Jet Propulsion Laboratory, California Institute of Technology, under contract with the National Aeronautics and Space Administration. Finally, the National Radio Astronomy Observatory is a facility of the National Science Foundation operated under cooperative agreement by Associated Universities, Inc.

\end{acknowledgements}

\bibliography{master_ref}{}

\begin{thebibliography}{}
\expandafter\ifx\csname natexlab\endcsname\relax\def\natexlab#1{#1}\fi
\providecommand{\url}[1]{\href{#1}{#1}}
\providecommand{\dodoi}[1]{doi:~\href{http://doi.org/#1}{\nolinkurl{#1}}}
\providecommand{\doeprint}[1]{\href{http://ascl.net/#1}{\nolinkurl{http://ascl.net/#1}}}
\providecommand{\doarXiv}[1]{\href{https://arxiv.org/abs/#1}{\nolinkurl{https://arxiv.org/abs/#1}}}

\bibitem[{{Adamo} {et~al.}(2017){Adamo}, {Ryon}, {Messa}, {Kim}, {Grasha},
  {Cook}, {Calzetti}, {Lee}, {Whitmore}, {Elmegreen}, {Ubeda}, {Smith},
  {Bright}, {Runnholm}, {Andrews}, {Fumagalli}, {Gouliermis}, {Kahre}, {Nair},
  {Thilker}, {Walterbos}, {Wofford}, {Aloisi}, {Ashworth}, {Brown}, {Chandar},
  {Christian}, {Cignoni}, {Clayton}, {Dale}, {de Mink}, {Dobbs}, {Elmegreen},
  {Evans}, {Gallagher}, {Grebel}, {Herrero}, {Hunter}, {Johnson}, {Kennicutt},
  {Krumholz}, {Lennon}, {Levay}, {Martin}, {Nota}, {{\"O}stlin}, {Pellerin},
  {Prieto}, {Regan}, {Sabbi}, {Sacchi}, {Schaerer}, {Schiminovich}, {Shabani},
  {Tosi}, {Van Dyk}, \& {Zackrisson}}]{aa17}
{Adamo}, A., {Ryon}, J.~E., {Messa}, M., {et~al.} 2017, \apj, 841, 131,
  \dodoi{10.3847/1538-4357/aa7132}

\bibitem[{{Adamo} {et~al.}(2020){Adamo}, {Hollyhead}, {Messa}, {Ryon}, {Bajaj},
  {Runnholm}, {Aalto}, {Calzetti}, {Gallagher}, {Hayes}, {Kruijssen},
  {K{\"o}nig}, {Larsen}, {Melinder}, {Sabbi}, {Smith}, \& {{\"O}stlin}}]{aa20a}
{Adamo}, A., {Hollyhead}, K., {Messa}, M., {et~al.} 2020, \mnras, 499, 3267,
  \dodoi{10.1093/mnras/staa2380}

\bibitem[{{Ali} {et~al.}(2022){Ali}, {Bending}, \& {Dobbs}}]{ali22}
{Ali}, A.~A., {Bending}, T. J.~R., \& {Dobbs}, C.~L. 2022, \mnras, 510, 5592,
  \dodoi{10.1093/mnras/stac025}

\bibitem[{{Allamandola} {et~al.}(1989){Allamandola}, {Tielens}, \&
  {Barker}}]{allamandola98}
{Allamandola}, L.~J., {Tielens}, A.~G.~G.~M., \& {Barker}, J.~R. 1989, \apjs,
  71, 733, \dodoi{10.1086/191396}

\bibitem[{{Andersen} {et~al.}(2009){Andersen}, {Zinnecker}, {Moneti},
  {McCaughrean}, {Brandl}, {Brandner}, {Meylan}, \& {Hunter}}]{andersen09}
{Andersen}, M., {Zinnecker}, H., {Moneti}, A., {et~al.} 2009, \apj, 707, 1347,
  \dodoi{10.1088/0004-637X/707/2/1347}

\bibitem[{{Barnes} {et~al.}(2021){Barnes}, {Glover}, {Kreckel}, {Ostriker},
  {Bigiel}, {Belfiore}, {Be{\v{s}}li{\'c}}, {Blanc}, {Chevance}, {Dale},
  {Egorov}, {Eibensteiner}, {Emsellem}, {Grasha}, {Groves}, {Klessen},
  {Kruijssen}, {Leroy}, {Longmore}, {Lopez}, {McElroy}, {Meidt}, {Murphy},
  {Rosolowsky}, {Saito}, {Santoro}, {Schinnerer}, {Schruba}, {Sun}, {Watkins},
  \& {Williams}}]{barnes21}
{Barnes}, A.~T., {Glover}, S.~C.~O., {Kreckel}, K., {et~al.} 2021, \mnras, 508,
  5362, \dodoi{10.1093/mnras/stab2958}

\bibitem[{{Bending} {et~al.}(2022){Bending}, {Dobbs}, \& {Bate}}]{bending22}
{Bending}, T. J.~R., {Dobbs}, C.~L., \& {Bate}, M.~R. 2022, \mnras, 513, 2088,
  \dodoi{10.1093/mnras/stac965}

\bibitem[{{Bertelli} {et~al.}(1994){Bertelli}, {Bressan}, {Chiosi}, {Fagotto},
  \& {Nasi}}]{padova94}
{Bertelli}, G., {Bressan}, A., {Chiosi}, C., {Fagotto}, F., \& {Nasi}, E. 1994,
  \aaps, 106, 275

\bibitem[{{Bertin} \& {Arnouts}(1996)}]{sex}
{Bertin}, E., \& {Arnouts}, S. 1996, \aaps, 117, 393,
  \dodoi{10.1051/aas:1996164}

\bibitem[{{Bik} {et~al.}(2006){Bik}, {Kaper}, \& {Waters}}]{bik06}
{Bik}, A., {Kaper}, L., \& {Waters}, L.~B.~F.~M. 2006, \aap, 455, 561,
  \dodoi{10.1051/0004-6361:20042403}

\bibitem[{{Bik} {et~al.}(2012){Bik}, {Henning}, {Stolte}, {Brandner},
  {Gouliermis}, {Gennaro}, {Pasquali}, {Rochau}, {Beuther}, {Ageorges},
  {Seifert}, {Wang}, \& {Kudryavtseva}}]{bik12}
{Bik}, A., {Henning}, T., {Stolte}, A., {et~al.} 2012, \apj, 744, 87,
  \dodoi{10.1088/0004-637X/744/2/87}

\bibitem[{{Black} \& {van Dishoeck}(1987)}]{black87}
{Black}, J.~H., \& {van Dishoeck}, E.~F. 1987, \apj, 322, 412,
  \dodoi{10.1086/165740}

\bibitem[{{B{\"o}ker} {et~al.}(2008){B{\"o}ker}, {Falc{\'o}n-Barroso},
  {Schinnerer}, {Knapen}, \& {Ryder}}]{boker08}
{B{\"o}ker}, T., {Falc{\'o}n-Barroso}, J., {Schinnerer}, E., {Knapen}, J.~H.,
  \& {Ryder}, S. 2008, \aj, 135, 479, \dodoi{10.1088/0004-6256/135/2/479}

\bibitem[{{B{\"o}ker} {et~al.}(2022){B{\"o}ker}, {Arribas}, {L{\"u}tzgendorf},
  {Alves de Oliveira}, {Beck}, {Birkmann}, {Bunker}, {Charlot}, {de Marchi},
  {Ferruit}, {Giardino}, {Jakobsen}, {Kumari}, {L{\'o}pez-Caniego}, {Maiolino},
  {Manjavacas}, {Marston}, {Moseley}, {Muzerolle}, {Ogle}, {Pirzkal},
  {Rauscher}, {Rawle}, {Rix}, {Sabbi}, {Sargent}, {Sirianni}, {te Plate},
  {Valenti}, {Willott}, \& {Zeidler}}]{boker22}
{B{\"o}ker}, T., {Arribas}, S., {L{\"u}tzgendorf}, N., {et~al.} 2022, \aap,
  661, A82, \dodoi{10.1051/0004-6361/202142589}

\bibitem[{{Bonne} {et~al.}(2023){Bonne}, {Kabanovic}, {Schneider}, {Zavagno},
  {Keilmann}, {Simon}, {Buchbender}, {G{\"u}sten}, {Jacob}, {Jacobs}, {Kavak},
  {Polles}, {Tiwari}, {Wyrowski}, \& {Tielens}}]{bonne23}
{Bonne}, L., {Kabanovic}, S., {Schneider}, N., {et~al.} 2023, \aap, 679, L5,
  \dodoi{10.1051/0004-6361/202347721}

\bibitem[{{Brunetti} {et~al.}(2021){Brunetti}, {Wilson}, {Sliwa}, {Schinnerer},
  {Aalto}, \& {Peck}}]{brunetti21}
{Brunetti}, N., {Wilson}, C.~D., {Sliwa}, K., {et~al.} 2021, \mnras, 500, 4730,
  \dodoi{10.1093/mnras/staa3425}

\bibitem[{{Brunetti} {et~al.}(2024){Brunetti}, {Wilson}, {He}, {Sun}, {Leroy},
  {Rosolowsky}, {Bemis}, {Bigiel}, {Groves}, {Saito}, \&
  {Schinnerer}}]{brunetti24}
{Brunetti}, N., {Wilson}, C.~D., {He}, H., {et~al.} 2024, \mnras, 530, 597,
  \dodoi{10.1093/mnras/stae890}

\bibitem[{{Bruzual} \& {Charlot}(2003)}]{bc03}
{Bruzual}, G., \& {Charlot}, S. 2003, \mnras, 344, 1000,
  \dodoi{10.1046/j.1365-8711.2003.06897.x}

\bibitem[{{Chandar} {et~al.}(2010){Chandar}, {Whitmore}, {Kim}, {Kaleida},
  {Mutchler}, {Calzetti}, {Saha}, {O'Connell}, {Balick}, {Bond}, {Carollo},
  {Disney}, {Dopita}, {Frogel}, {Hall}, {Holtzman}, {Kimble}, {McCarthy},
  {Paresce}, {Silk}, {Trauger}, {Walker}, {Windhorst}, \& {Young}}]{rc10c}
{Chandar}, R., {Whitmore}, B.~C., {Kim}, H., {et~al.} 2010, \apj, 719, 966,
  \dodoi{10.1088/0004-637X/719/1/966}

\bibitem[{{Chevance} {et~al.}(2020){Chevance}, {Kruijssen}, {Krumholz},
  {Groves}, {Keller}, {Hughes}, {Glover}, {Henshaw}, {Herrera}, {Kim}, {Leroy},
  {Pety}, {Razza}, {Rosolowsky}, {Schinnerer}, {Schruba}, {Barnes}, {Bigiel},
  {Blanc}, {Emsellem}, {Faesi}, {Grasha}, {Klessen}, {Kreckel}, {Liu},
  {Longmore}, {Meidt}, {Querejeta}, {Saito}, {Sun}, \& {Usero}}]{chevance20a}
{Chevance}, M., {Kruijssen}, J.~M.~D., {Krumholz}, M.~R., {et~al.} 2020, arXiv
  e-prints, arXiv:2010.13788.
\newblock \doarXiv{2010.13788}

\bibitem[{{Chevance} {et~al.}(2022){Chevance}, {Kruijssen}, {Krumholz},
  {Groves}, {Keller}, {Hughes}, {Glover}, {Henshaw}, {Herrera}, {Kim}, {Leroy},
  {Pety}, {Razza}, {Rosolowsky}, {Schinnerer}, {Schruba}, {Barnes}, {Bigiel},
  {Blanc}, {Dale}, {Emsellem}, {Faesi}, {Grasha}, {Klessen}, {Kreckel}, {Liu},
  {Longmore}, {Meidt}, {Querejeta}, {Saito}, {Sun}, \& {Usero}}]{chevance22}
---. 2022, \mnras, 509, 272, \dodoi{10.1093/mnras/stab2938}

\bibitem[{{Cook} {et~al.}(2019){Cook}, {Lee}, {Adamo}, {Kim}, {Chandar},
  {Whitmore}, {Mok}, {Ryon}, {Dale}, {Calzetti}, {Andrews}, {Aloisi},
  {Ashworth}, {Bright}, {Brown}, {Christian}, {Cignoni}, {Clayton}, {da Silva},
  {de Mink}, {Dobbs}, {Elmegreen}, {Elmegreen}, {Evans}, {Fumagalli},
  {Gallagher}, {Gouliermis}, {Grasha}, {Grebel}, {Herrero}, {Hunter}, {Jensen},
  {Johnson}, {Kahre}, {Kennicutt}, {Krumholz}, {Lee}, {Lennon}, {Linden},
  {Martin}, {Messa}, {Nair}, {Nota}, {{\"O}stlin}, {Parziale}, {Pellerin},
  {Regan}, {Sabbi}, {Sacchi}, {Schaerer}, {Schiminovich}, {Shabani}, {Slane},
  {Small}, {Smith}, {Smith}, {Taibi}, {Thilker}, {de la Torre}, {Tosi},
  {Turner}, {Ubeda}, {Van Dyk}, {Walterbos}, \& {Wofford}}]{cook19}
{Cook}, D.~O., {Lee}, J.~C., {Adamo}, A., {et~al.} 2019, \mnras, 484, 4897,
  \dodoi{10.1093/mnras/stz331}

\bibitem[{{Cresci} {et~al.}(2010){Cresci}, {Vanzi}, {Sauvage}, {Santangelo}, \&
  {van der Werf}}]{cresci10}
{Cresci}, G., {Vanzi}, L., {Sauvage}, M., {Santangelo}, G., \& {van der Werf},
  P. 2010, \aap, 520, A82, \dodoi{10.1051/0004-6361/201014864}

\bibitem[{{Dale} {et~al.}(2023){Dale}, {Boquien}, {Barnes}, {Belfiore},
  {Bigiel}, {Cao}, {Chandar}, {Chastenet}, {Chevance}, {Deger}, {Egorov},
  {Grasha}, {Groves}, {Hassani}, {Henny}, {Klessen}, {Kreckel}, {Kruijssen},
  {Larson}, {Lee}, {Leroy}, {Liu}, {Murphy}, {Rosolowsky}, {Sandstrom},
  {Schinnerer}, {Sutter}, {Thilker}, {Watkins}, {Whitmore}, \&
  {Williams}}]{dale23}
{Dale}, D.~A., {Boquien}, M., {Barnes}, A.~T., {et~al.} 2023, \apjl, 944, L23,
  \dodoi{10.3847/2041-8213/aca769}

\bibitem[{{Dale} {et~al.}(2012){Dale}, {Ercolano}, \& {Bonnell}}]{dale12}
{Dale}, J.~E., {Ercolano}, B., \& {Bonnell}, I.~A. 2012, \mnras, 424, 377,
  \dodoi{10.1111/j.1365-2966.2012.21205.x}

\bibitem[{{Dale} {et~al.}(2014){Dale}, {Ngoumou}, {Ercolano}, \&
  {Bonnell}}]{dale14}
{Dale}, J.~E., {Ngoumou}, J., {Ercolano}, B., \& {Bonnell}, I.~A. 2014, \mnras,
  442, 694, \dodoi{10.1093/mnras/stu816}

\bibitem[{{Della Bruna} {et~al.}(2022){Della Bruna}, {Adamo}, {McLeod},
  {Smith}, {Savard}, {Robert}, {Sun}, {Amram}, {Bik}, {Blair}, {Long},
  {Renaud}, {Walterbos}, \& {Usher}}]{della-bruna22}
{Della Bruna}, L., {Adamo}, A., {McLeod}, A.~F., {et~al.} 2022, \aap, 666, A29,
  \dodoi{10.1051/0004-6361/202243395}

\bibitem[{{Doherty} {et~al.}(1995){Doherty}, {Puxley}, {Lumsden}, \&
  {Doyon}}]{doherty95}
{Doherty}, R.~M., {Puxley}, P.~J., {Lumsden}, S.~L., \& {Doyon}, R. 1995,
  \mnras, 277, 577, \dodoi{10.1093/mnras/277.2.577}

\bibitem[{{Donnan} {et~al.}(2024){Donnan}, {Garc{\'\i}a-Bernete}, {Rigopoulou},
  {Pereira-Santaella}, {Roche}, \& {Alonso-Herrero}}]{donnan24}
{Donnan}, F.~R., {Garc{\'\i}a-Bernete}, I., {Rigopoulou}, D., {et~al.} 2024,
  \mnras, 529, 1386, \dodoi{10.1093/mnras/stae612}

\bibitem[{{Draine} \& {Li}(2007)}]{dl07}
{Draine}, B.~T., \& {Li}, A. 2007, \apj, 657, 810, \dodoi{10.1086/511055}

\bibitem[{{Draine} {et~al.}(2021){Draine}, {Li}, {Hensley}, {Hunt},
  {Sandstrom}, \& {Smith}}]{dl21}
{Draine}, B.~T., {Li}, A., {Hensley}, B.~S., {et~al.} 2021, \apj, 917, 3,
  \dodoi{10.3847/1538-4357/abff51}

\bibitem[{{Egorov} {et~al.}(2023){Egorov}, {Kreckel}, {Sandstrom}, {Leroy},
  {Glover}, {Groves}, {Kruijssen}, {Barnes}, {Belfiore}, {Bigiel}, {Blanc},
  {Boquien}, {Cao}, {Chastenet}, {Chevance}, {Congiu}, {Dale}, {Emsellem},
  {Grasha}, {Klessen}, {Larson}, {Liu}, {Murphy}, {Pan}, {Pessa}, {Pety},
  {Rosolowsky}, {Scheuermann}, {Schinnerer}, {Sutter}, {Thilker}, {Watkins}, \&
  {Williams}}]{egorov23}
{Egorov}, O.~V., {Kreckel}, K., {Sandstrom}, K.~M., {et~al.} 2023, \apjl, 944,
  L16, \dodoi{10.3847/2041-8213/acac92}

\bibitem[{{Ferland} {et~al.}(2017){Ferland}, {Chatzikos}, {Guzm{\'a}n},
  {Lykins}, {van Hoof}, {Williams}, {Abel}, {Badnell}, {Keenan}, {Porter}, \&
  {Stancil}}]{cloudy17}
{Ferland}, G.~J., {Chatzikos}, M., {Guzm{\'a}n}, F., {et~al.} 2017, \rmxaa, 53,
  385.
\newblock \doarXiv{1705.10877}

\bibitem[{{Fitzpatrick}(1999)}]{fitz99}
{Fitzpatrick}, E.~L. 1999, \pasp, 111, 63, \dodoi{10.1086/316293}

\bibitem[{{F{\"o}rster Schreiber} {et~al.}(2001){F{\"o}rster Schreiber},
  {Genzel}, {Lutz}, {Kunze}, \& {Sternberg}}]{schreiber01}
{F{\"o}rster Schreiber}, N.~M., {Genzel}, R., {Lutz}, D., {Kunze}, D., \&
  {Sternberg}, A. 2001, \apj, 552, 544, \dodoi{10.1086/320546}

\bibitem[{{Fukushima} {et~al.}(2020){Fukushima}, {Yajima}, {Sugimura},
  {Hosokawa}, {Omukai}, \& {Matsumoto}}]{fukushima20}
{Fukushima}, H., {Yajima}, H., {Sugimura}, K., {et~al.} 2020, \mnras, 497,
  3830, \dodoi{10.1093/mnras/staa2062}

\bibitem[{{Gaia Collaboration} {et~al.}(2018){Gaia Collaboration}, {Mignard},
  {Klioner}, {Lindegren}, {Hern{\'a}ndez}, {Bastian}, {Bombrun}, {Hobbs},
  {Lammers}, {Michalik}, {Ramos-Lerate}, {Biermann},
  {Fern{\'a}ndez-Hern{\'a}ndez}, {Geyer}, {Hilger}, {Siddiqui},
  {Steidelm{\"u}ller}, {Babusiaux}, {Barache}, {Lambert}, {Andrei}, {Bourda},
  {Charlot}, {Brown}, {Vallenari}, {Prusti}, {de Bruijne}, {Bailer-Jones},
  {Evans}, {Eyer}, {Jansen}, {Jordi}, {Luri}, {Panem}, {Pourbaix}, {Randich},
  {Sartoretti}, {Soubiran}, {van Leeuwen}, {Walton}, {Arenou}, {Cropper},
  {Drimmel}, {Katz}, {Lattanzi}, {Bakker}, {Cacciari}, {Casta{\~n}eda},
  {Chaoul}, {Cheek}, {De Angeli}, {Fabricius}, {Guerra}, {Holl}, {Masana},
  {Messineo}, {Mowlavi}, {Nienartowicz}, {Panuzzo}, {Portell}, {Riello},
  {Seabroke}, {Tanga}, {Th{\'e}venin}, {Gracia-Abril}, {Comoretto},
  {Garcia-Reinaldos}, {Teyssier}, {Altmann}, {Andrae}, {Audard},
  {Bellas-Velidis}, {Benson}, {Berthier}, {Blomme}, {Burgess}, {Busso},
  {Carry}, {Cellino}, {Clementini}, {Clotet}, {Creevey}, {Davidson}, {De
  Ridder}, {Delchambre}, {Dell'Oro}, {Ducourant}, {Fouesneau}, {Fr{\'e}mat},
  {Galluccio}, {Garc{\'\i}a-Torres}, {Gonz{\'a}lez-N{\'u}{\~n}ez},
  {Gonz{\'a}lez-Vidal}, {Gosset}, {Guy}, {Halbwachs}, {Hambly}, {Harrison},
  {Hestroffer}, {Hodgkin}, {Hutton}, {Jasniewicz}, {Jean-Antoine-Piccolo},
  {Jordan}, {Korn}, {Krone-Martins}, {Lanzafame}, {Lebzelter}, {L{\"o}ffler},
  {Manteiga}, {Marrese}, {Mart{\'\i}n-Fleitas}, {Moitinho}, {Mora}, {Muinonen},
  {Osinde}, {Pancino}, {Pauwels}, {Petit}, {Recio-Blanco}, {Richards},
  {Rimoldini}, {Robin}, {Sarro}, {Siopis}, {Smith}, {Sozzetti}, {S{\"u}veges},
  {Torra}, {van Reeven}, {Abbas}, {Abreu Aramburu}, {Accart}, {Aerts},
  {Altavilla}, {{\'A}lvarez}, {Alvarez}, {Alves}, {Anderson}, {Anglada Varela},
  {Antiche}, {Antoja}, {Arcay}, {Astraatmadja}, {Bach}, {Baker},
  {Balaguer-N{\'u}{\~n}ez}, {Balm}, {Barata}, {Barbato}, {Barblan}, {Barklem},
  {Barrado}, {Barros}, {Barstow}, {Bartholom{\'e} Mu{\~n}oz}, {Bassilana},
  {Becciani}, {Bellazzini}, {Berihuete}, {Bertone}, {Bianchi}, {Bienaym{\'e}},
  {Blanco-Cuaresma}, {Boch}, {Boeche}, {Borrachero}, {Bossini}, {Bouquillon},
  {Bragaglia}, {Bramante}, {Breddels}, {Bressan}, {Brouillet},
  {Br{\"u}semeister}, {Brugaletta}, {Bucciarelli}, {Burlacu}, {Busonero},
  {Butkevich}, {Buzzi}, {Caffau}, {Cancelliere}, {Cannizzaro}, {Cantat-Gaudin},
  {Carballo}, {Carlucci}, {Carrasco}, {Casamiquela}, {Castellani},
  {Castro-Ginard}, {Chemin}, {Chiavassa}, {Cocozza}, {Costigan}, {Cowell},
  {Crifo}, {Crosta}, {Crowley}, {Cuypers}, {Dafonte}, {Damerdji}, {Dapergolas},
  {David}, {David}, {de Laverny}, {De Luise}, {De March}, {de Souza}, {de
  Torres}, {Debosscher}, {del Pozo}, {Delbo}, {Delgado}, {Delgado}, {Diakite},
  {Diener}, {Distefano}, {Dolding}, {Drazinos}, {Dur{\'a}n}, {Edvardsson},
  {Enke}, {Eriksson}, {Esquej}, {Eynard Bontemps}, {Fabre}, {Fabrizio},
  {Faigler}, {Falc{\~a}o}, {Farr{\`a}s Casas}, {Federici}, {Fedorets},
  {Fernique}, {Figueras}, {Filippi}, {Findeisen}, {Fonti}, {Fraile}, {Fraser},
  {Fr{\'e}zouls}, {Gai}, {Galleti}, {Garabato}, {Garc{\'\i}a-Sedano},
  {Garofalo}, {Garralda}, {Gavel}, {Gavras}, {Gerssen}, {Giacobbe}, {Gilmore},
  {Girona}, {Giuffrida}, {Glass}, {Gomes}, {Granvik}, {Gueguen}, {Guerrier},
  {Guiraud}, {Guti{\'e}}, {Haigron}, {Hatzidimitriou}, {Hauser}, {Haywood},
  {Heiter}, {Helmi}, {Heu}, {Hofmann}, {Holland}, {Huckle}, {Hypki}, {Icardi},
  {Jan{\ss}en}, {Jevardat de Fombelle}, {Jonker}, {Juh{\'a}sz}, {Julbe},
  {Karampelas}, {Kewley}, {Klar}, {Kochoska}, {Kohley}, {Kolenberg},
  {Kontizas}, {Kontizas}, {Koposov}, {Kordopatis}, {Kostrzewa-Rutkowska},
  {Koubsky}, {Lanza}, {Lasne}, {Lavigne}, {Le Fustec}, {Le Poncin-Lafitte},
  {Lebreton}, {Leccia}, {Leclerc}, {Lecoeur-Taibi}, {Lenhardt}, {Leroux},
  {Liao}, {Licata}, {Lindstr{\o}m}, {Lister}, {Livanou}, {Lobel}, {L{\'o}pez},
  {Managau}, {Mann}, {Mantelet}, {Marchal}, {Marchant}, {Marconi}, {Marinoni},
  {Marschalk{\'o}}, {Marshall}, {Martino}, {Marton}, {Mary}, {Massari},
  {Matijevi{\v{c}}}, {Mazeh}, {McMillan}, {Messina}, {Millar}, {Molina},
  {Molinaro}, {Moln{\'a}r}, {Montegriffo}, {Mor}, {Morbidelli}, {Morel},
  {Morris}, {Mulone}, {Muraveva}, {Musella}, {Nelemans}, {Nicastro}, {Noval},
  {O'Mullane}, {Ord{\'e}novic}, {Ord{\'o}{\~n}ez-Blanco}, {Osborne}, {Pagani},
  {Pagano}, {Pailler}, {Palacin}, {Palaversa}, {Panahi}, {Pawlak},
  {Piersimoni}, {Pineau}, {Plachy}, {Plum}, {Poggio}, {Poujoulet},
  {Pr{\v{s}}a}, {Pulone}, {Racero}, {Ragaini}, {Rambaux}, {Regibo},
  {Reyl{\'e}}, {Riclet}, {Ripepi}, {Riva}, {Rivard}, {Rixon}, {Roegiers},
  {Roelens}, {Romero-G{\'o}mez}, {Rowell}, {Royer}, {Ruiz-Dern}, {Sadowski},
  {Sagrist{\`a} Sell{\'e}s}, {Sahlmann}, {Salgado}, {Salguero}, {Sanna},
  {Santana-Ros}, {Sarasso}, {Savietto}, {Schultheis}, {Sciacca}, {Segol},
  {Segovia}, {S{\'e}gransan}, {Shih}, {Siltala}, {Silva}, {Smart}, {Smith},
  {Solano}, {Solitro}, {Sordo}, {Soria Nieto}, {Souchay}, {Spagna}, {Spoto},
  {Stampa}, {Steele}, {Stephenson}, {Stoev}, {Suess}, {Surdej}, {Szabados},
  {Szegedi-Elek}, {Tapiador}, {Taris}, {Tauran}, {Taylor}, {Teixeira},
  {Terrett}, {Teyssandier}, {Thuillot}, {Titarenko}, {Torra Clotet}, {Turon},
  {Ulla}, {Utrilla}, {Uzzi}, {Vaillant}, {Valentini}, {Valette}, {van Elteren},
  {Van Hemelryck}, {van Leeuwen}, {Vaschetto}, {Vecchiato}, {Veljanoski},
  {Viala}, {Vicente}, {Vogt}, {von Essen}, {Voss}, {Votruba}, {Voutsinas},
  {Walmsley}, {Weiler}, {Wertz}, {Wevers}, {Wyrzykowski}, {Yoldas},
  {{\v{Z}}erjal}, {Ziaeepour}, {Zorec}, {Zschocke}, {Zucker}, {Zurbach}, \&
  {Zwitter}}]{gaiadr2}
{Gaia Collaboration}, {Mignard}, F., {Klioner}, S.~A., {et~al.} 2018, \aap,
  616, A14, \dodoi{10.1051/0004-6361/201832916}

\bibitem[{{Galliano} {et~al.}(2008){Galliano}, {Alloin}, {Pantin}, {Granato},
  {Delva}, {Silva}, {Lagage}, \& {Panuzzo}}]{galliano08}
{Galliano}, E., {Alloin}, D., {Pantin}, E., {et~al.} 2008, \aap, 492, 3,
  \dodoi{10.1051/0004-6361:20077621}

\bibitem[{{Garc{\'\i}a-Mar{\'\i}n} {et~al.}(2006){Garc{\'\i}a-Mar{\'\i}n},
  {Colina}, {Arribas}, {Alonso-Herrero}, \& {Mediavilla}}]{garcia-marin06}
{Garc{\'\i}a-Mar{\'\i}n}, M., {Colina}, L., {Arribas}, S., {Alonso-Herrero},
  A., \& {Mediavilla}, E. 2006, \apj, 650, 850, \dodoi{10.1086/507411}

\bibitem[{{Gazak} {et~al.}(2013){Gazak}, {Bastian}, {Kudritzki}, {Adamo},
  {Davies}, {Plez}, \& {Urbaneja}}]{gazak13}
{Gazak}, J.~Z., {Bastian}, N., {Kudritzki}, R.~P., {et~al.} 2013, \mnras, 430,
  L35, \dodoi{10.1093/mnrasl/sls043}

\bibitem[{{Geen} {et~al.}(2021){Geen}, {Bieri}, {Rosdahl}, \& {de
  Koter}}]{geen21}
{Geen}, S., {Bieri}, R., {Rosdahl}, J., \& {de Koter}, A. 2021, \mnras, 501,
  1352, \dodoi{10.1093/mnras/staa3705}

\bibitem[{{Geen} {et~al.}(2018){Geen}, {Watson}, {Rosdahl}, {Bieri}, {Klessen},
  \& {Hennebelle}}]{geen18}
{Geen}, S., {Watson}, S.~K., {Rosdahl}, J., {et~al.} 2018, \mnras, 481, 2548,
  \dodoi{10.1093/mnras/sty2439}

\bibitem[{{Goddard} {et~al.}(2010){Goddard}, {Bastian}, \&
  {Kennicutt}}]{goddard10}
{Goddard}, Q.~E., {Bastian}, N., \& {Kennicutt}, R.~C. 2010, \mnras, 405, 857,
  \dodoi{10.1111/j.1365-2966.2010.16511.x}

\bibitem[{{Gordon} {et~al.}(2022){Gordon}, {Bohlin}, {Sloan}, {Rieke}, {Volk},
  {Boyer}, {Muzerolle}, {Schlawin}, {Deustua}, {Hines}, {Kraemer}, {Mullally},
  \& {Su}}]{gordon22}
{Gordon}, K.~D., {Bohlin}, R., {Sloan}, G.~C., {et~al.} 2022, \aj, 163, 267,
  \dodoi{10.3847/1538-3881/ac66dc}

\bibitem[{{Grasha} {et~al.}(2018){Grasha}, {Calzetti}, {Bittle}, {Johnson},
  {Donovan Meyer}, {Kennicutt}, {Elmegreen}, {Adamo}, {Krumholz}, {Fumagalli},
  {Grebel}, {Gouliermis}, {Cook}, {Gallagher}, {Aloisi}, {Dale}, {Linden},
  {Sacchi}, {Thilker}, {Walterbos}, {Messa}, {Wofford}, \& {Smith}}]{grasha18}
{Grasha}, K., {Calzetti}, D., {Bittle}, L., {et~al.} 2018, \mnras, 481, 1016,
  \dodoi{10.1093/mnras/sty2154}

\bibitem[{{Grasha} {et~al.}(2019){Grasha}, {Calzetti}, {Adamo}, {Kennicutt},
  {Elmegreen}, {Messa}, {Dale}, {Fedorenko}, {Mahadevan}, {Grebel},
  {Fumagalli}, {Kim}, {Dobbs}, {Gouliermis}, {Ashworth}, {Gallagher}, {Smith},
  {Tosi}, {Whitmore}, {Schinnerer}, {Colombo}, {Hughes}, {Leroy}, \&
  {Meidt}}]{grasha19}
{Grasha}, K., {Calzetti}, D., {Adamo}, A., {et~al.} 2019, \mnras, 483, 4707,
  \dodoi{10.1093/mnras/sty3424}

\bibitem[{{Guszejnov} {et~al.}(2022){Guszejnov}, {Grudi{\'c}}, {Offner},
  {Faucher-Gigu{\`e}re}, {Hopkins}, \& {Rosen}}]{guszejnov22}
{Guszejnov}, D., {Grudi{\'c}}, M.~Y., {Offner}, S. S.~R., {et~al.} 2022,
  \mnras, 515, 4929, \dodoi{10.1093/mnras/stac2060}

\bibitem[{{Habart} {et~al.}(2011){Habart}, {Abergel}, {Boulanger}, {Joblin},
  {Verstraete}, {Compi{\`e}gne}, {Pineau Des For{\^e}ts}, \& {Le
  Bourlot}}]{habart11}
{Habart}, E., {Abergel}, A., {Boulanger}, F., {et~al.} 2011, \aap, 527, A122,
  \dodoi{10.1051/0004-6361/20077327}

\bibitem[{{Hannon} {et~al.}(2019){Hannon}, {Lee}, {Whitmore}, {Chandar},
  {Adamo}, {Mobasher}, {Aloisi}, {Calzetti}, {Cignoni}, {Cook}, {Dale},
  {Deger}, {Della Bruna}, {Elmegreen}, {Gouliermis}, {Grasha}, {Grebel},
  {Herrero}, {Hunter}, {Johnson}, {Kennicutt}, {Kim}, {Sacchi}, {Smith},
  {Thilker}, {Turner}, {Walterbos}, \& {Wofford}}]{hannon19}
{Hannon}, S., {Lee}, J.~C., {Whitmore}, B.~C., {et~al.} 2019, \mnras, 490,
  4648, \dodoi{10.1093/mnras/stz2820}

\bibitem[{{Hannon} {et~al.}(2022){Hannon}, {Lee}, {Whitmore}, {Mobasher},
  {Thilker}, {Chandar}, {Adamo}, {Wofford}, {Orozco-Duarte}, {Calzetti}, {Della
  Bruna}, {Kreckel}, {Groves}, {Barnes}, {Boquien}, {Belfiore}, \&
  {Linden}}]{hannon22}
---. 2022, \mnras, 512, 1294, \dodoi{10.1093/mnras/stac550}

\bibitem[{{Hanson} {et~al.}(2002){Hanson}, {Luhman}, \& {Rieke}}]{hanson02}
{Hanson}, M.~M., {Luhman}, K.~L., \& {Rieke}, G.~H. 2002, \apjs, 138, 35,
  \dodoi{10.1086/324073}

\bibitem[{{Herrera} \& {Boulanger}(2017)}]{herrera17}
{Herrera}, C.~N., \& {Boulanger}, F. 2017, \aap, 600, A139,
  \dodoi{10.1051/0004-6361/201628454}

\bibitem[{{Hinshaw} {et~al.}(2013){Hinshaw}, {Larson}, {Komatsu}, {Spergel},
  {Bennett}, {Dunkley}, {Nolta}, {Halpern}, {Hill}, {Odegard}, {Page}, {Smith},
  {Weiland}, {Gold}, {Jarosik}, {Kogut}, {Limon}, {Meyer}, {Tucker}, {Wollack},
  \& {Wright}}]{wmap}
{Hinshaw}, G., {Larson}, D., {Komatsu}, E., {et~al.} 2013, \apjs, 208, 19,
  \dodoi{10.1088/0067-0049/208/2/19}

\bibitem[{{Hollenbach} \& {Tielens}(1999)}]{hat99}
{Hollenbach}, D.~J., \& {Tielens}, A.~G.~G.~M. 1999, Reviews of Modern Physics,
  71, 173, \dodoi{10.1103/RevModPhys.71.173}

\bibitem[{{Hollyhead} {et~al.}(2015){Hollyhead}, {Bastian}, {Adamo},
  {Silva-Villa}, {Dale}, {Ryon}, \& {Gazak}}]{hollyhead15}
{Hollyhead}, K., {Bastian}, N., {Adamo}, A., {et~al.} 2015, \mnras, 449, 1106,
  \dodoi{10.1093/mnras/stv331}

\bibitem[{{Howard} {et~al.}(2017){Howard}, {Pudritz}, \& {Harris}}]{howard17}
{Howard}, C.~S., {Pudritz}, R.~E., \& {Harris}, W.~E. 2017, \mnras, 470, 3346,
  \dodoi{10.1093/mnras/stx1363}

\bibitem[{{Jakobsen} {et~al.}(2022){Jakobsen}, {Ferruit}, {Alves de Oliveira},
  {Arribas}, {Bagnasco}, {Barho}, {Beck}, {Birkmann}, {B{\"o}ker}, {Bunker},
  {Charlot}, {de Jong}, {de Marchi}, {Ehrenwinkler}, {Falcolini}, {Fels},
  {Franx}, {Franz}, {Funke}, {Giardino}, {Gnata}, {Holota}, {Honnen}, {Jensen},
  {Jentsch}, {Johnson}, {Jollet}, {Karl}, {Kling}, {K{\"o}hler}, {Kolm},
  {Kumari}, {Lander}, {Lemke}, {L{\'o}pez-Caniego}, {L{\"u}tzgendorf},
  {Maiolino}, {Manjavacas}, {Marston}, {Maschmann}, {Maurer}, {Messerschmidt},
  {Moseley}, {Mosner}, {Mott}, {Muzerolle}, {Pirzkal}, {Pittet}, {Plitzke},
  {Posselt}, {Rapp}, {Rauscher}, {Rawle}, {Rix}, {R{\"o}del}, {Rumler},
  {Sabbi}, {Salvignol}, {Schmid}, {Sirianni}, {Smith}, {Strada}, {te Plate},
  {Valenti}, {Wettemann}, {Wiehe}, {Wiesmayer}, {Willott}, {Wright}, {Zeidler},
  \& {Zincke}}]{jakobsen22}
{Jakobsen}, P., {Ferruit}, P., {Alves de Oliveira}, C., {et~al.} 2022, \aap,
  661, A80, \dodoi{10.1051/0004-6361/202142663}

\bibitem[{{Kankare} {et~al.}(2018{\natexlab{a}}){Kankare}, {Mattila}, {Ryder},
  {Kool}, {Kotak}, {Kotilainen}, {Kuncarayakti}, {Perez-Torres},
  {Randriamanakoto}, {Reynolds}, {Romero-Canizales}, {Smartt}, \&
  {Vaisanen}}]{kankare18a}
{Kankare}, E., {Mattila}, S., {Ryder}, S., {et~al.} 2018{\natexlab{a}}, The
  Astronomer's Telegram, 11156, 1

\bibitem[{{Kankare} {et~al.}(2018{\natexlab{b}}){Kankare}, {Taubenberger},
  {Vogl}, {Floers}, {Malesani}, {Rubin}, {Inserra}, {Maguire}, {Smartt},
  {Yaron}, {Young}, {Kool}, {Kotak}, {Kotilainen}, {Kuncarayakti}, {Mattila},
  {Perez-Torres}, {Randriamanakoto}, {Reynolds}, {Romero-Canizales}, {Ryder},
  \& {Vaisanen}}]{kankare18b}
{Kankare}, E., {Taubenberger}, S., {Vogl}, C., {et~al.} 2018{\natexlab{b}}, The
  Astronomer's Telegram, 11778, 1

\bibitem[{{Kim} {et~al.}(2023){Kim}, {Chevance}, {Kruijssen}, {Barnes},
  {Bigiel}, {Blanc}, {Boquien}, {Cao}, {Congiu}, {Dale}, {Egorov}, {Faesi},
  {Glover}, {Grasha}, {Groves}, {Hassani}, {Hughes}, {Klessen}, {Kreckel},
  {Larson}, {Lee}, {Leroy}, {Liu}, {Longmore}, {Meidt}, {Pan}, {Pety},
  {Querejeta}, {Rosolowsky}, {Saito}, {Sandstrom}, {Schinnerer}, {Smith},
  {Usero}, {Watkins}, \& {Williams}}]{kim23}
{Kim}, J., {Chevance}, M., {Kruijssen}, J.~M.~D., {et~al.} 2023, \apjl, 944,
  L20, \dodoi{10.3847/2041-8213/aca90a}

\bibitem[{{Kim} {et~al.}(2018){Kim}, {Kim}, \& {Ostriker}}]{kim18}
{Kim}, J.-G., {Kim}, W.-T., \& {Ostriker}, E.~C. 2018, \apj, 859, 68,
  \dodoi{10.3847/1538-4357/aabe27}

\bibitem[{{Krause} {et~al.}(2013){Krause}, {Fierlinger}, {Diehl}, {Burkert},
  {Voss}, \& {Ziegler}}]{krause13}
{Krause}, M., {Fierlinger}, K., {Diehl}, R., {et~al.} 2013, \aap, 550, A49,
  \dodoi{10.1051/0004-6361/201220060}

\bibitem[{{Kruijssen} {et~al.}(2019){Kruijssen}, {Schruba}, {Chevance},
  {Longmore}, {Hygate}, {Haydon}, {McLeod}, {Dalcanton}, {Tacconi}, \& {van
  Dishoeck}}]{kruijssen19}
{Kruijssen}, J.~M.~D., {Schruba}, A., {Chevance}, M., {et~al.} 2019, \nat, 569,
  519, \dodoi{10.1038/s41586-019-1194-3}

\bibitem[{{Krumholz} {et~al.}(2019){Krumholz}, {McKee}, \&
  {Bland-Hawthorn}}]{krumholz19}
{Krumholz}, M.~R., {McKee}, C.~F., \& {Bland-Hawthorn}, J. 2019, \araa, 57,
  227, \dodoi{10.1146/annurev-astro-091918-104430}

\bibitem[{{Lai} {et~al.}(2020){Lai}, {Smith}, {Baba}, {Spoon}, \&
  {Imanishi}}]{lai20}
{Lai}, T. S.~Y., {Smith}, J.~D.~T., {Baba}, S., {Spoon}, H. W.~W., \&
  {Imanishi}, M. 2020, \apj, 905, 55, \dodoi{10.3847/1538-4357/abc002}

\bibitem[{{Lai} {et~al.}(2022){Lai}, {Armus}, {U}, {D{\'\i}az-Santos},
  {Larson}, {Evans}, {Malkan}, {Appleton}, {Rich}, {M{\"u}ller-S{\'a}nchez},
  {Inami}, {Bohn}, {McKinney}, {Finnerty}, {Law}, {Linden}, {Medling},
  {Privon}, {Song}, {Stierwalt}, {van der Werf}, {Barcos-Mu{\~n}oz}, {Smith},
  {Togi}, {Aalto}, {B{\"o}ker}, {Charmandaris}, {Howell}, {Iwasawa}, {Kemper},
  {Mazzarella}, {Murphy}, {Brown}, {Hayward}, {Marshall}, {Sanders}, \&
  {Surace}}]{lai22}
{Lai}, T. S.~Y., {Armus}, L., {U}, V., {et~al.} 2022, \apjl, 941, L36,
  \dodoi{10.3847/2041-8213/ac9ebf}

\bibitem[{{Lai} {et~al.}(2023){Lai}, {Armus}, {Bianchin}, {D{\'\i}az-Santos},
  {Linden}, {Privon}, {Inami}, {U}, {Bohn}, {Evans}, {Larson}, {Hensley},
  {Smith}, {Malkan}, {Song}, {Stierwalt}, {van der Werf}, {McKinney}, {Aalto},
  {Buiten}, {Rich}, {Charmandaris}, {Appleton}, {Barcos-Mu{\~n}oz},
  {B{\"o}ker}, {Finnerty}, {Kader}, {Law}, {Medling}, {Brown}, {Hayward},
  {Howell}, {Iwasawa}, {Kemper}, {Marshall}, {Mazzarella},
  {M{\"u}ller-S{\'a}nchez}, {Murphy}, {Sanders}, \& {Surace}}]{lai23}
{Lai}, T. S.~Y., {Armus}, L., {Bianchin}, M., {et~al.} 2023, \apjl, 957, L26,
  \dodoi{10.3847/2041-8213/ad0387}

\bibitem[{{Larsen} \& {Richtler}(2000)}]{larsen00}
{Larsen}, S.~S., \& {Richtler}, T. 2000, \aap, 354, 836.
\newblock \doarXiv{astro-ph/0001198}

\bibitem[{{Larsen} {et~al.}(2011){Larsen}, {de Mink}, {Eldridge}, {Langer},
  {Bastian}, {Seth}, {Smith}, {Brodie}, \& {Efremov}}]{larsen11}
{Larsen}, S.~S., {de Mink}, S.~E., {Eldridge}, J.~J., {et~al.} 2011, \aap, 532,
  A147, \dodoi{10.1051/0004-6361/201117185}

\bibitem[{{Larson} {et~al.}(2020){Larson}, {D{\'\i}az-Santos}, {Armus},
  {Privon}, {Linden}, {Evans}, {Howell}, {Charmandaris}, {U}, {Sanders},
  {Stierwalt}, {Barcos-Mu{\~n}oz}, {Rich}, {Medling}, {Cook},
  {Oklop{\^{c}}i{\'c}}, {Murphy}, \& {Bonfini}}]{klarson20}
{Larson}, K.~L., {D{\'\i}az-Santos}, T., {Armus}, L., {et~al.} 2020, \apj, 888,
  92, \dodoi{10.3847/1538-4357/ab5dc3}

\bibitem[{{Leger} {et~al.}(1989){Leger}, {D'Hendecourt}, \&
  {Defourneau}}]{leger89}
{Leger}, A., {D'Hendecourt}, L., \& {Defourneau}, D. 1989, \aap, 216, 148

\bibitem[{{Leitherer} {et~al.}(2018){Leitherer}, {Byler}, {Lee}, \&
  {Levesque}}]{leitherer18}
{Leitherer}, C., {Byler}, N., {Lee}, J.~C., \& {Levesque}, E.~M. 2018, \apj,
  865, 55, \dodoi{10.3847/1538-4357/aada84}

\bibitem[{{Leitherer} {et~al.}(2014){Leitherer}, {Ekstr{\"o}m}, {Meynet},
  {Schaerer}, {Agienko}, \& {Levesque}}]{leitherer14}
{Leitherer}, C., {Ekstr{\"o}m}, S., {Meynet}, G., {et~al.} 2014, \apjs, 212,
  14, \dodoi{10.1088/0067-0049/212/1/14}

\bibitem[{{Leitherer} {et~al.}(1999){Leitherer}, {Schaerer}, {Goldader},
  {Gonz{\'a}lez Delgado}, {Robert}, {Kune}, {de Mello}, {Devost}, \&
  {Heckman}}]{sb99}
{Leitherer}, C., {Schaerer}, D., {Goldader}, J.~D., {et~al.} 1999, \apjs, 123,
  3, \dodoi{10.1086/313233}

\bibitem[{{Levy} {et~al.}(2021){Levy}, {Bolatto}, {Leroy}, {Emig}, {Gorski},
  {Krieger}, {Lenki{\'c}}, {Meier}, {Mills}, {Ott}, {Rosolowsky}, {Tarantino},
  {Veilleux}, {Walter}, {Wei{\ss}}, \& {Zwaan}}]{levy21}
{Levy}, R.~C., {Bolatto}, A.~D., {Leroy}, A.~K., {et~al.} 2021, \apj, 912, 4,
  \dodoi{10.3847/1538-4357/abec84}

\bibitem[{{Lin} {et~al.}(2020){Lin}, {Calzetti}, {Kong}, {Adamo}, {Cignoni},
  {Cook}, {Dale}, {Grasha}, {Grebel}, {Messa}, {Sacchi}, \& {Smith}}]{lin20}
{Lin}, Z., {Calzetti}, D., {Kong}, X., {et~al.} 2020, \apj, 896, 16,
  \dodoi{10.3847/1538-4357/ab9106}

\bibitem[{{Linden} {et~al.}(2021){Linden}, {Evans}, {Larson}, {Privon},
  {Armus}, {Rich}, {Diaz-Santos}, {Murphy}, {Song}, {Barcos-Munoz}, {Howell},
  {Charmandaris}, {Inami}, {U}, {Surace}, {Mazzarella}, \&
  {Calzetti}}]{linden21}
{Linden}, S., {Evans}, A., {Larson}, K., {et~al.} 2021, arXiv e-prints,
  arXiv:2110.03638.
\newblock \doarXiv{2110.03638}

\bibitem[{{Linden} {et~al.}(2017){Linden}, {Evans}, {Rich}, {Larson}, {Armus},
  {D{\'\i}az-Santos}, {Privon}, {Howell}, {Inami}, {Kim}, {Chien}, {Vavilkin},
  {Mazzarella}, {Modica}, {Surace}, {Manning}, {Abdullah}, {Blake}, {Yarber},
  \& {Lambert}}]{stl17}
{Linden}, S.~T., {Evans}, A.~S., {Rich}, J., {et~al.} 2017, \apj, 843, 91,
  \dodoi{10.3847/1538-4357/aa7266}

\bibitem[{{Linden} {et~al.}(2023){Linden}, {Evans}, {Armus}, {Rich}, {Larson},
  {Lai}, {Privon}, {U}, {Inami}, {Bohn}, {Song}, {Barcos-Mu{\~n}oz},
  {Charmandaris}, {Medling}, {Stierwalt}, {Diaz-Santos}, {B{\"o}ker}, {van der
  Werf}, {Aalto}, {Appleton}, {Brown}, {Hayward}, {Howell}, {Iwasawa},
  {Kemper}, {Frayer}, {Law}, {Malkan}, {Marshall}, {Mazzarella}, {Murphy},
  {Sanders}, \& {Surace}}]{linden23}
{Linden}, S.~T., {Evans}, A.~S., {Armus}, L., {et~al.} 2023, \apjl, 944, L55,
  \dodoi{10.3847/2041-8213/acb335}

\bibitem[{{Lopez} {et~al.}(2011){Lopez}, {Krumholz}, {Bolatto}, {Prochaska}, \&
  {Ramirez-Ruiz}}]{lopez11}
{Lopez}, L.~A., {Krumholz}, M., {Bolatto}, A., {Prochaska}, J.~X., \&
  {Ramirez-Ruiz}, E. 2011, in American Astronomical Society Meeting Abstracts,
  Vol. 217, American Astronomical Society Meeting Abstracts \#217, 132.04

\bibitem[{{Lopez} {et~al.}(2014){Lopez}, {Krumholz}, {Bolatto}, {Prochaska},
  {Ramirez-Ruiz}, \& {Castro}}]{lopez14}
{Lopez}, L.~A., {Krumholz}, M.~R., {Bolatto}, A.~D., {et~al.} 2014, \apj, 795,
  121, \dodoi{10.1088/0004-637X/795/2/121}

\bibitem[{{Lucas} {et~al.}(2020){Lucas}, {Bonnell}, \& {Dale}}]{lucas20}
{Lucas}, W.~E., {Bonnell}, I.~A., \& {Dale}, J.~E. 2020, \mnras, 493, 4700,
  \dodoi{10.1093/mnras/staa451}

\bibitem[{{Lumsden} {et~al.}(2001){Lumsden}, {Puxley}, \& {Hoare}}]{lumsden01}
{Lumsden}, S.~L., {Puxley}, P.~J., \& {Hoare}, M.~G. 2001, \mnras, 320, 83,
  \dodoi{10.1046/j.1365-8711.2001.03954.x}

\bibitem[{{Lumsden} {et~al.}(2003){Lumsden}, {Puxley}, {Hoare}, {Moore}, \&
  {Ridge}}]{lumsden03}
{Lumsden}, S.~L., {Puxley}, P.~J., {Hoare}, M.~G., {Moore}, T.~J.~T., \&
  {Ridge}, N.~A. 2003, \mnras, 340, 799,
  \dodoi{10.1046/j.1365-8711.2003.06337.x}

\bibitem[{{Mart{\'\i}n-Hern{\'a}ndez}
  {et~al.}(2008){Mart{\'\i}n-Hern{\'a}ndez}, {Bik}, {Puga}, {N{\"u}rnberger},
  \& {Bronfman}}]{mn08}
{Mart{\'\i}n-Hern{\'a}ndez}, N.~L., {Bik}, A., {Puga}, E., {N{\"u}rnberger},
  D.~E.~A., \& {Bronfman}, L. 2008, \aap, 489, 229,
  \dodoi{10.1051/0004-6361:200810336}

\bibitem[{{Maschmann} {et~al.}(2024){Maschmann}, {Lee}, {Thilker}, {Whitmore},
  {Deger}, {Boquien}, {Chandar}, {Dale}, {Wofford}, {Hannon}, {Larson},
  {Leroy}, {Schinnerer}, {Rosolowsky}, {Ubeda}, {Barnes}, {Emsellem}, {Grasha},
  {Groves}, {Kim}, {Klessen}, {Kreckel}, {Levy}, {Pinna}, {Rodriguez}, {Tian},
  \& {Williams}}]{maschmann24}
{Maschmann}, D., {Lee}, J.~C., {Thilker}, D.~A., {et~al.} 2024, arXiv e-prints,
  arXiv:2403.04901, \dodoi{10.48550/arXiv.2403.04901}

\bibitem[{{Matthews} {et~al.}(2018){Matthews}, {Johnson}, {Whitmore}, {Brogan},
  {Leroy}, \& {Indebetouw}}]{matthews18}
{Matthews}, A.~M., {Johnson}, K.~E., {Whitmore}, B.~C., {et~al.} 2018, \apj,
  862, 147, \dodoi{10.3847/1538-4357/aac958}

\bibitem[{{Messa} {et~al.}(2018){Messa}, {Adamo}, {{\"O}stlin}, {Calzetti},
  {Grasha}, {Grebel}, {Shabani}, {Chandar}, {Dale}, {Dobbs}, {Elmegreen},
  {Fumagalli}, {Gouliermis}, {Kim}, {Smith}, {Thilker}, {Tosi}, {Ubeda},
  {Walterbos}, {Whitmore}, {Fedorenko}, {Mahadevan}, {Andrews}, {Bright},
  {Cook}, {Kahre}, {Nair}, {Pellerin}, {Ryon}, {Ahmad}, {Beale}, {Brown},
  {Clarkson}, {Guidarelli}, {Parziale}, {Turner}, \& {Weber}}]{mm18a}
{Messa}, M., {Adamo}, A., {{\"O}stlin}, G., {et~al.} 2018, \mnras, 473, 996,
  \dodoi{10.1093/mnras/stx2403}

\bibitem[{{Micelotta} {et~al.}(2010){Micelotta}, {Jones}, \&
  {Tielens}}]{micelotta10}
{Micelotta}, E.~R., {Jones}, A.~P., \& {Tielens}, A.~G.~G.~M. 2010, \aap, 510,
  A36, \dodoi{10.1051/0004-6361/200911682}

\bibitem[{{Murray} {et~al.}(2010){Murray}, {Quataert}, \&
  {Thompson}}]{murray10}
{Murray}, N., {Quataert}, E., \& {Thompson}, T.~A. 2010, \apj, 709, 191,
  \dodoi{10.1088/0004-637X/709/1/191}

\bibitem[{{Pellegrini} {et~al.}(2011){Pellegrini}, {Baldwin}, \&
  {Ferland}}]{pellegrini11}
{Pellegrini}, E.~W., {Baldwin}, J.~A., \& {Ferland}, G.~J. 2011, \apj, 738, 34,
  \dodoi{10.1088/0004-637X/738/1/34}

\bibitem[{{Perrin} {et~al.}(2014){Perrin}, {Sivaramakrishnan}, {Lajoie},
  {Elliott}, {Pueyo}, {Ravindranath}, \& {Albert}}]{perrin14}
{Perrin}, M.~D., {Sivaramakrishnan}, A., {Lajoie}, C.-P., {et~al.} 2014, in
  Space Telescopes and Instrumentation 2014: Optical, Infrared, and Millimeter
  Wave, Vol. 9143, 91433X, \dodoi{10.1117/12.2056689}

\bibitem[{{Rich} {et~al.}(2012){Rich}, {Torrey}, {Kewley}, {Dopita}, \&
  {Rupke}}]{jrich12}
{Rich}, J.~A., {Torrey}, P., {Kewley}, L.~J., {Dopita}, M.~A., \& {Rupke},
  D.~S.~N. 2012, \apj, 753, 5, \dodoi{10.1088/0004-637X/753/1/5}

\bibitem[{{Rieke} {et~al.}(2023){Rieke}, {Kelly}, {Misselt}, {Stansberry},
  {Boyer}, {Beatty}, {Egami}, {Florian}, {Greene}, {Hainline}, {Leisenring},
  {Roellig}, {Schlawin}, {Sun}, {Tinnin}, {Williams}, {Willmer}, {Wilson},
  {Clark}, {Rohrbach}, {Brooks}, {Canipe}, {Correnti}, {DiFelice}, {Gennaro},
  {Girard}, {Hartig}, {Hilbert}, {Koekemoer}, {Nikolov}, {Pirzkal}, {Rest},
  {Robberto}, {Sunnquist}, {Telfer}, {Wu}, {Ferry}, {Lewis}, {Baum},
  {Beichman}, {Doyon}, {Dressler}, {Eisenstein}, {Ferrarese}, {Hodapp},
  {Horner}, {Jaffe}, {Johnstone}, {Krist}, {Martin}, {McCarthy}, {Meyer},
  {Rieke}, {Trauger}, \& {Young}}]{nircam2}
{Rieke}, M.~J., {Kelly}, D.~M., {Misselt}, K., {et~al.} 2023, \pasp, 135,
  028001, \dodoi{10.1088/1538-3873/acac53}

\bibitem[{{Rodriguez} {et~al.}(2022){Rodriguez}, {Lee}, {Whitmore}, {Thilker},
  {Maschmann}, {Chandar}, {Dale}, {Kruijssen}, {Boquien}, {Grasha}, {Watkins},
  {Barnes}, {Sormani}, {Williams}, {Kim}, {Anand}, {Chevance}, {Bigiel},
  {Leroy}, {Klessen}, {Rosolowsky}, {Sandstrom}, {Hassani}, {Kim},
  {Schinnerer}, {Larson}, {Deger}, {Liu}, {Faesi}, {Cao}, {Belfiore}, {Pessa},
  {Kreckel}, {Groves}, {Pety}, {Indebetouw}, {Egorov}, {Blanc}, {Saito},
  {Emsellem}, \& {Hughes}}]{rodriguez22}
{Rodriguez}, J., {Lee}, J., {Whitmore}, B., {et~al.} 2022, arXiv e-prints,
  arXiv:2211.13426, \dodoi{10.48550/arXiv.2211.13426}

\bibitem[{{Ryon} {et~al.}(2017){Ryon}, {Gallagher}, {Smith}, {Adamo},
  {Calzetti}, {Bright}, {Cignoni}, {Cook}, {Dale}, {Elmegreen}, {Fumagalli},
  {Gouliermis}, {Grasha}, {Grebel}, {Kim}, {Messa}, {Thilker}, \&
  {Ubeda}}]{ryon17}
{Ryon}, J.~E., {Gallagher}, J.~S., {Smith}, L.~J., {et~al.} 2017, \apj, 841,
  92, \dodoi{10.3847/1538-4357/aa719e}

\bibitem[{{S{\'a}nchez-Garc{\'\i}a} {et~al.}(2022){S{\'a}nchez-Garc{\'\i}a},
  {Pereira-Santaella}, {Garc{\'\i}a-Burillo}, {Colina}, {Alonso-Herrero},
  {Villar-Mart{\'\i}n}, {Saito}, {D{\'\i}az-Santos}, {Piqueras L{\'o}pez},
  {Arribas}, {Bellocchi}, {Cazzoli}, \& {Labiano}}]{sg22}
{S{\'a}nchez-Garc{\'\i}a}, M., {Pereira-Santaella}, M., {Garc{\'\i}a-Burillo},
  S., {et~al.} 2022, \aap, 659, A102, \dodoi{10.1051/0004-6361/202141963}

\bibitem[{{Sandstrom} {et~al.}(2023){Sandstrom}, {Chastenet}, {Sutter},
  {Leroy}, {Egorov}, {Williams}, {Bolatto}, {Boquien}, {Cao}, {Dale}, {Lee},
  {Rosolowsky}, {Schinnerer}, {Barnes}, {Belfiore}, {Bigiel}, {Chevance},
  {Grasha}, {Groves}, {Hassani}, {Hughes}, {Klessen}, {Kruijssen}, {Larson},
  {Liu}, {Lopez}, {Meidt}, {Murphy}, {Sormani}, {Thilker}, \&
  {Watkins}}]{sandstrom23}
{Sandstrom}, K.~M., {Chastenet}, J., {Sutter}, J., {et~al.} 2023, \apjl, 944,
  L7, \dodoi{10.3847/2041-8213/acb0cf}

\bibitem[{{Schlafly} \& {Finkbeiner}(2011)}]{schlafly11}
{Schlafly}, E.~F., \& {Finkbeiner}, D.~P. 2011, \apj, 737, 103,
  \dodoi{10.1088/0004-637X/737/2/103}

\bibitem[{{Seok} {et~al.}(2014){Seok}, {Hirashita}, \& {Asano}}]{seok14}
{Seok}, J.~Y., {Hirashita}, H., \& {Asano}, R.~S. 2014, \mnras, 439, 2186,
  \dodoi{10.1093/mnras/stu120}

\bibitem[{{Speagle} {et~al.}(2014){Speagle}, {Steinhardt}, {Capak}, \&
  {Silverman}}]{speagle14}
{Speagle}, J.~S., {Steinhardt}, C.~L., {Capak}, P.~L., \& {Silverman}, J.~D.
  2014, \apjs, 214, 15, \dodoi{10.1088/0067-0049/214/2/15}

\bibitem[{{Teh} {et~al.}(2023){Teh}, {Grasha}, {Krumholz}, {Battisti},
  {Calzetti}, {Rousseau-Nepton}, {Rhea}, {Adamo}, {Kennicutt}, {Grebel},
  {Cook}, {Combes}, {Messa}, {Linden}, {Klessen}, {Vilchez}, {Fumagalli},
  {McLeod}, {Smith}, {Chemin}, {Wang}, {Sabbi}, {Sacchi}, {Petric}, {Della
  Bruna}, \& {Boselli}}]{teh23}
{Teh}, J.~W., {Grasha}, K., {Krumholz}, M.~R., {et~al.} 2023, \mnras, 524,
  1191, \dodoi{10.1093/mnras/stad1780}

\bibitem[{{Tielens}(2008)}]{tielens08}
{Tielens}, A.~G.~G.~M. 2008, \araa, 46, 289,
  \dodoi{10.1146/annurev.astro.46.060407.145211}

\bibitem[{{Trancho} {et~al.}(2007){Trancho}, {Bastian}, {Miller}, \&
  {Schweizer}}]{trancho07}
{Trancho}, G., {Bastian}, N., {Miller}, B.~W., \& {Schweizer}, F. 2007, \apj,
  664, 284, \dodoi{10.1086/518886}

\bibitem[{{Westmoquette} {et~al.}(2014){Westmoquette}, {Bastian}, {Smith},
  {Seth}, {Gallagher}, {O'Connell}, {Ryon}, {Silich}, {Mayya},
  {Mu{\~n}oz-Tu{\~n}{\'o}n}, \& {Rosa Gonz{\'a}lez}}]{westmonquette14}
{Westmoquette}, M.~S., {Bastian}, N., {Smith}, L.~J., {et~al.} 2014, \apj, 789,
  94, \dodoi{10.1088/0004-637X/789/2/94}

\bibitem[{{Whelan} {et~al.}(2011){Whelan}, {Johnson}, {Whitney}, {Indebetouw},
  \& {Wood}}]{whelan11}
{Whelan}, D.~G., {Johnson}, K.~E., {Whitney}, B.~A., {Indebetouw}, R., \&
  {Wood}, K. 2011, \apj, 729, 111, \dodoi{10.1088/0004-637X/729/2/111}

\bibitem[{{Whitmore} \& {Zhang}(2002)}]{bcm02b}
{Whitmore}, B.~C., \& {Zhang}, Q. 2002, \aj, 124, 1418, \dodoi{10.1086/341822}

\bibitem[{{Whitmore} {et~al.}(2010){Whitmore}, {Chandar}, {Schweizer},
  {Rothberg}, {Leitherer}, {Rieke}, {Rieke}, {Blair}, {Mengel}, \&
  {Alonso-Herrero}}]{bcm10}
{Whitmore}, B.~C., {Chandar}, R., {Schweizer}, F., {et~al.} 2010, \aj, 140, 75,
  \dodoi{10.1088/0004-6256/140/1/75}

\bibitem[{{Whitmore} {et~al.}(2011){Whitmore}, {Chandar}, {Kim}, {Kaleida},
  {Mutchler}, {Stankiewicz}, {Calzetti}, {Saha}, {O'Connell}, {Balick}, {Bond},
  {Carollo}, {Disney}, {Dopita}, {Frogel}, {Hall}, {Holtzman}, {Kimble},
  {McCarthy}, {Paresce}, {Silk}, {Trauger}, {Walker}, {Windhorst}, \&
  {Young}}]{bcm11}
{Whitmore}, B.~C., {Chandar}, R., {Kim}, H., {et~al.} 2011, \apj, 729, 78,
  \dodoi{10.1088/0004-637X/729/2/78}

\bibitem[{{Whitmore} {et~al.}(2023){Whitmore}, {Chandar}, {Rodr{\'\i}guez},
  {Lee}, {Emsellem}, {Floyd}, {Kim}, {Kruijssen}, {Mok}, {Sormani}, {Boquien},
  {Dale}, {Faesi}, {Henny}, {Hannon}, {Thilker}, {White}, {Barnes}, {Bigiel},
  {Chevance}, {Henshaw}, {Klessen}, {Leroy}, {Liu}, {Maschmann}, {Meidt},
  {Rosolowsky}, {Schinnerer}, {Sun}, {Watkins}, \& {Williams}}]{whitmore23}
{Whitmore}, B.~C., {Chandar}, R., {Rodr{\'\i}guez}, M.~J., {et~al.} 2023,
  \apjl, 944, L14, \dodoi{10.3847/2041-8213/acae94}

\bibitem[{{Willmer}(2018)}]{willmer18}
{Willmer}, C. N.~A. 2018, \apjs, 236, 47, \dodoi{10.3847/1538-4365/aabfdf}

\bibitem[{{Wu} {et~al.}(2019){Wu}, {Baker}, {Heckman}, {Hicks}, {Lutz}, \&
  {Tacconi}}]{wu19}
{Wu}, J.~F., {Baker}, A.~J., {Heckman}, T.~M., {et~al.} 2019, \apj, 887, 251,
  \dodoi{10.3847/1538-4357/ab5953}

\bibitem[{{Zackrisson} {et~al.}(2011){Zackrisson}, {Rydberg}, {Schaerer},
  {{\"O}stlin}, \& {Tuli}}]{yggdrasil}
{Zackrisson}, E., {Rydberg}, C.-E., {Schaerer}, D., {{\"O}stlin}, G., \&
  {Tuli}, M. 2011, \apj, 740, 13, \dodoi{10.1088/0004-637X/740/1/13}

\end{thebibliography}
\bibliographystyle{aasjournal}

\begin{deluxetable*}{lll|ccccccc}
\center
\tablecaption{NIRCam/NIRSpec Nuclear Cluster Catalog\label{tbl-1}}
\tabletypesize{\footnotesize}
\tablewidth{0pt}
\tablehead{
\colhead{ID} & \colhead{RA\tablenotemark{a}}  & \colhead{Dec} & \colhead{F150W-F200W} & \colhead{F200W-F335M}  & \colhead{Age (Myr)} &  \colhead{Mass ($M_{\odot}$)\tablenotemark{b}} & \colhead{EW (3.3$\mu$m)\tablenotemark{c}} & \colhead{E(B - V)\tablenotemark{d}} & \colhead{$\chi^{2}$}}
\startdata
R1 & 156.96338 & -43.905387 & 1.327 & 1.007 & 2.5 & 4.18 & 0.28 & 5.75 & 7.1 \\
R2 & 156.96353 & -43.905470 & 0.403 & -0.1028 & 6.3 & 8.772 & 0.14 & 2.0 & 1.5 \\
R3 & 156.96349 & -43.905067 & 0.3185 & -0.1277 & 6.0 & 3.55 & 0.11 & 0.75 & 1.3 \\
R4 & 156.96401 & -43.904206 & 0.1176 & -0.4568 & - & 9.831 & 0.052 & 0.25 & 1.3 \\
R5 & 156.96394 & -43.903470 & 0.1611 & -0.7704 & 6.7 & 22.8 & -0.024 & 0.75 & 1.4 \\
R9 & 156.96326 & -43.904067 & 0.8953 & 0.4948 & 1.0 & 6.134 & 0.31 & 3.5 & 7.9 \\
R10 & 156.96311 & -43.904123 & 0.7439 & 0.4491 & 3.1 & 3.306 & 0.36 & 3.0 & 3.6  \\
R13 & 156.96405 & -43.903896 & 0.4199 & 0.1587 & 3.8 & 4.155 & 0.37 & 1.5 & 3.3 \\
\enddata
\tablenotetext{a}{The RA and DEC are given in the J2000 coordinate system.}
\tablenotetext{b}{The stellar mass of eahc YMC is given as $M_{\odot}$x$10^{5}$}
\tablenotetext{c}{The 3.3$\mu$m PAH equivalent width is measured in $\mu$m after a local background subtraction has been applied.}
\tablenotetext{d}{The color excess E(B - V) is given in unites of mag.}
\end{deluxetable*}

\end{document}